\documentclass[final,3p,times]{elsarticle}

\usepackage{amsmath,amssymb,amsfonts}
\usepackage{graphicx}
\usepackage{textcomp}
\usepackage{xcolor}
\usepackage{hyperref}
\usepackage{booktabs} 
\usepackage{multirow}
\usepackage{threeparttablex}
\usepackage{ragged2e}
\usepackage{algorithm}
\usepackage{algpseudocode}
\usepackage{subcaption}
\usepackage[table]{xcolor}

\journal{Biomedical Signal Processing and Control}

\begin{document}

\begin{frontmatter}

\title{Non-Invasive Reconstruction of Intracranial EEG Across the Deep Temporal Lobe from Scalp EEG based on Conditional Normalizing Flow}

\author[1]{Dongyi He} % 12223050207@stu.cqut.edu.cn
\author[1]{Bin Jiang\corref{cor1}} % jb20200132@cqut.edu.cn
\author[1]{Kecheng Feng} % 18225402538@stu.cqut.edu.cn
\author[1]{Luyin Zhang} % 2790793903@qq.com
\author[1]{Ling Liu} % newlyliuling@stu.cqut.edu.cn
\author[1]{Yuxuan Li} % 2366923468@qq.com
\author[2]{Yun Zhao} % zyyy90@163.com
\author[1]{He Yan} % yanhe@cqut.edu.cn

\affiliation[1]{organization={School of Artificial Intelligence, Chongqing University of Technology},
      city={Chongqing},
      postcode={400054}, 
      country={China}}

\affiliation[2]{organization={School of Smart Health, Chongqing Polytechnic University of Electronic Technology},
      city={Chongqing},
      postcode={401331}, 
      country={China}}

\cortext[cor1]{Corresponding author. Email: jb20200132@cqut.edu.cn}

\begin{abstract}
Although obtaining deep brain activity from non-invasive scalp electroencephalography (EEG) is crucial for neuroscience and clinical diagnosis, directly generating high-fidelity intracranial electroencephalography (iEEG) signals remains a largely unexplored field, limiting our understanding of deep brain dynamics. Current research primarily focuses on traditional signal processing or source localization methods, which struggle to capture the complex waveforms and random characteristics of iEEG. To address this critical challenge, this paper introduces NeuroFlowNet, a novel cross-modal generative framework that reconstructs iEEG signals from scalp EEG over multiple medial temporal lobe (MTL) subregions covered by depth microelectrodes in a public synchronized EEG-iEEG dataset. NeuroFlowNet is built on Conditional Normalizing Flow (CNF), which models complex conditional probability distributions through reversible transformations and latent-variable sampling. Additionally, the model integrates a multi-scale architecture and self-attention mechanisms to robustly capture fine-grained temporal details and long-range dependencies. Validation on a publicly available synchronized EEG-iEEG dataset under both cross-trial and cross-session within-subject protocols shows that NeuroFlowNet can generate band-limited iEEG signals that closely match the ground truth in the time domain, reproduce low-/mid-frequency spectral characteristics (0.5-50~Hz, including the alpha band 8-13~Hz), and preserve the inter-channel correlation structure associated with MTL functional connectivity within the retained bandwidth. Importantly, the cross-session results remain comparable to the cross-trial results, indicating that the model retains stable reconstruction performance on unseen recording sessions from the same subject. Moreover, compared with representative deterministic regression baselines under the same subject-specific protocol, NeuroFlowNet yields consistently lower errors in the inter-channel correlation matrix, indicating better preservation of band-limited cross-channel dependency structure. A conditional sampling and uncertainty analysis further demonstrates non-degenerate stochastic outputs under repeated latent draws for the same EEG input, with predictive variance strongly associated with reconstruction error across validation segments. These results support the feasibility of non-invasive iEEG reconstruction over the recorded MTL subregions, offering a practical route toward studying deep-brain dynamics from scalp EEG under a subject-specific setting. The code of this study is available in \url{https://github.com/hdy6438/NeuroFlowNet}
\end{abstract}


\end{frontmatter}

\section{Introduction}
Electroencephalography (EEG) is a high-time-resolution brain activity monitoring technique that can track neuronal activation in real time and dynamically record brain activity. It plays a crucial role in brain-computer interfaces, clinical diagnosis, and cognitive neuroscience research~\cite{craik2019deep,kumari2022study,jiang2026comprehensive,li2025tale,he2026toward}. Among these, scalp electroencephalography (EEG) measures the distribution of potentials on the scalp resulting from the propagation of postsynaptic potentials and action potentials from neuronal clusters in the brain, as detected by electrodes placed on the scalp~\cite{michel2019eeg,li2025tale}. It offers the advantages of simplicity and high flexibility, making it widely used in clinical monitoring and diagnosis~\cite{biasiucci2018brain}. However, due to the influence of cranial volume conduction effects during signal propagation, its spatial resolution is relatively low~\cite{hassan2018electroencephalography,he2019electrophysiological,jiang2026comprehensive,zhao2025rgftslanet,he2025spec2volcamu}. That is, the measured signals represent the superimposed potentials formed by the activities of numerous sources within the brain on the scalp surface. These signals are characterized by weak potentials, susceptibility to external noise interference, and low signal-to-noise ratios, making it difficult to accurately decode the complex dynamic relationships of deep brain neural activities~\cite{ma2024attention,padfield2019eeg}. Recent studies have further explored neuroscience-informed dynamical EEG representations through spatiospectral modeling and neural manifold decoding, demonstrating the value of structured representation learning for neural decoding and feature transformation analysis~\cite{yu2025spatiospectral,yu2025neural}. This limitation restricts its further application in clinical settings. In contrast, intracranial electroencephalography (iEEG) can directly obtain neural activity signals from the cerebral cortex or deep nuclei, featuring high temporal-spatial resolution and high signal-to-noise ratio~\cite{parvizi2018human}. It is the gold standard for locating lesions in brain-related diseases and achieving high-precision neural decoding~\cite{parvizi2018promises,mukamel2012human}. However, its high surgical costs and the risks of immune reactions and infections following electrode implantation significantly limit its practical application~\cite{parvizi2018promises}.

Given the limited spatial resolution of EEG and the difficulty in obtaining iEEG signals, the brain electroencephalogram inverse problem has emerged as a research hotspot to achieve real-time monitoring and precise localization of neuronal activity~\cite{grech2008review,basri2024inverse,baillet2017magnetoencephalography}. Its core lies in how to infer the spatial distribution and temporal information of cortical and deep neuronal electrical activity from EEG signals recorded on the scalp~\cite{grech2008review,michel2019eeg}. By solving the EEG inverse problem, it is possible to map non-invasive EEG signals to neural signals of approximately iEEG quality~\cite{sohrabpour2015effect,hedrich2017comparison}. This not only helps to “unlock” dynamic information from deep brain regions under non-invasive conditions~\cite{he2018electrophysiological} but also provides new technical support for precise localization of epileptic foci, analysis of brain functional mechanisms, and neural rehabilitation interventions~\cite{seeber2019subcortical}.

However, significant differences in discharge patterns and signal characteristics between EEG and iEEG limit the effectiveness of traditional methods~\cite{ryvlin2025seeg,herff2020potential}. Parameter-based localization methods based on equivalent current dipoles (ECD) are constrained by the difficulty in accurately estimating the number of dipoles~\cite{michel2019eeg,veeramalla2020multiple}, while source reconstruction methods based on current distribution are hindered by discrepancies between prior constraints and actual neuronal activity~\cite{knosche2013prior}, both of which struggle to accurately reconstruct complex neural dynamics~\cite{ramirez2010neuroelectromagnetic,cui2019eeg}. Additionally, traditional machine learning methods rely on manual feature extraction~\cite{saeidi2021neural,hassan2023review,yu2019supervised} and have limited nonlinear modeling capabilities~\cite{saeidi2021neural}, further restricting their representation capabilities in EEG inversion.

In recent years, deep learning techniques have significantly improved EEG inversion performance through their powerful nonlinear fitting capabilities~\cite{rong2025deep,bozhkov2018overview,hui2022eeg}. For example, convolutional neural networks (CNNs) extract spatial correlations between multiple channels through spatial convolution, enhancing the accuracy of source space imaging~\cite{hecker2021convdip,huang2022electromagnetic,kaviri2024integrating}. Recurrent neural networks (RNNs) and their variants can model the temporal dynamic characteristics of signals, significantly improving the ability to distinguish between transient and sustained source activity~\cite{bore2021long,sikka2020investigating}. However, deep learning methods still face numerous challenges. Due to data limitations, model training often relies on large-scale synthetic data generated using finite element or boundary element models~\cite{doan2024scaling}. Synthetic data struggles to fully reflect the true complex mapping relationship between EEG and deep neuronal electrical activity, severely limiting the model's performance in real clinical applications. Additionally, the outputs of this method are primarily continuous neural activity distribution maps, which can locate positions but struggle to capture neural response waveforms, thereby impacting further decoding analysis~\cite{cisotto2024hveegnet}.

Generative models offer a new approach for high-fidelity mapping from EEG signals to iEEG signals, with the core mechanism being the use of generative networks to perform cross-modal conversion of easily obtainable non-invasive EEG signals into iEEG signals. Abdi-Sargezeh et al. were the first to introduce generative adversarial networks (GANs) to address this issue~\cite{abdi2023mapping}, successfully generating high-quality iEEG waveforms while demonstrating advantages in terms of efficiency and model complexity. Building on this, they proposed an EEG-to-EEG conversion model, named VAE-cGAN~\cite{abdi2025eeg}, combining variational autoencoders (VAE) and conditional generative adversarial networks (cGAN), and validated it on a dataset containing real iEEG signals recorded via Foramen Ovale electrodes. By encoding EEG into the latent space and regenerating iEEG signals, they further enhanced the fidelity and signal resolution of the mapped waveforms. The team achieved waveform reconstruction for the first time under conditions where real iEEG was used as a reference, providing a new technical pathway for non-invasive acquisition of high-fidelity iEEG signals. However, this research remains in its preliminary exploratory phase. Due to limitations such as the sampling range of Foramen Ovale channels and inherent defects like mode collapse in GAN models, current reconstruction is only feasible for localized regions of the medial temporal lobe (MTL), and expanding the number of reconstruction channels may significantly degrade reconstruction quality and effectiveness. Additionally, expanding the number of reconstructed channels may significantly degrade reconstruction quality and effectiveness. Additionally, due to the complexity and randomness of neural activity in the brain, models like GAN and VAE, which use deterministic mappings, often fail to adequately reflect the spatiotemporal diversity and randomness of iEEG signals. In addition to waveform fidelity, a fundamental requirement for iEEG reconstruction is to preserve the multi-channel dependency structure across intracranial contacts, since functional connectivity analyses are typically derived from inter-channel covariance/correlation patterns~\cite{chiarion2023connectivity,wei2025functional}. In an EEG-to-iEEG setting, the inverse mapping is underconstrained: multiple intracranial configurations can be compatible with similar scalp observations due to volume conduction and source mixing. Deterministic regressors trained with pointwise objectives may therefore converge to ``average'' solutions that reduce waveform error but attenuate network-level variability and distort the inter-channel correlation matrix.

To address these challenges, this study innovatively proposes a new cross-modal generative framework named NeuroFlowNet, which attempts for the first time to reconstruct iEEG signals across multiple MTL subregions using EEG. To capture the spatio-temporal diversity and randomness of iEEG signals, the model is based on Conditional Normalizing Flow (CNF), directly learning the complex conditional probability distribution $p(X_{iEEG}|X_{EEG})$ from EEG to iEEG. Compared to traditional deterministic mappings, the normalizing flow model maps complex data distributions to simple prior distributions through a series of reversible transformations, enabling repeated conditional sampling from latent Gaussian variables. This provides a direct mechanism for representing residual iEEG variability that is not uniquely determined by scalp EEG. Additionally, the model adopts a multi-scale architecture combined with self-attention mechanisms, further enhancing the fineness and robustness of iEEG waveform reconstruction. NeuroFlowNet has been validated for effectiveness on a publicly available dataset~\cite{boran2019persistent} of whole temporal lobe EEG-iEEG signals synchronously collected via Stereoelectroencephalography electrodes. Experimental results demonstrate that the model has advantages in signal fidelity, spectral feature reproduction, network structure recovery, and conditional sample uncertainty behavior, and that these properties remain stable when evaluation is shifted from unseen trials to unseen sessions within the same subject. We also assess computational latency under the final model configuration to examine whether the reversible generation process is compatible with real-time or near-real-time monitoring scenarios. This provides a more precise and reliable new paradigm for non-invasive analysis of deep brain dynamics.

\section{Methods}

\begin{figure*}
    \centering
    \includegraphics[width=\textwidth]{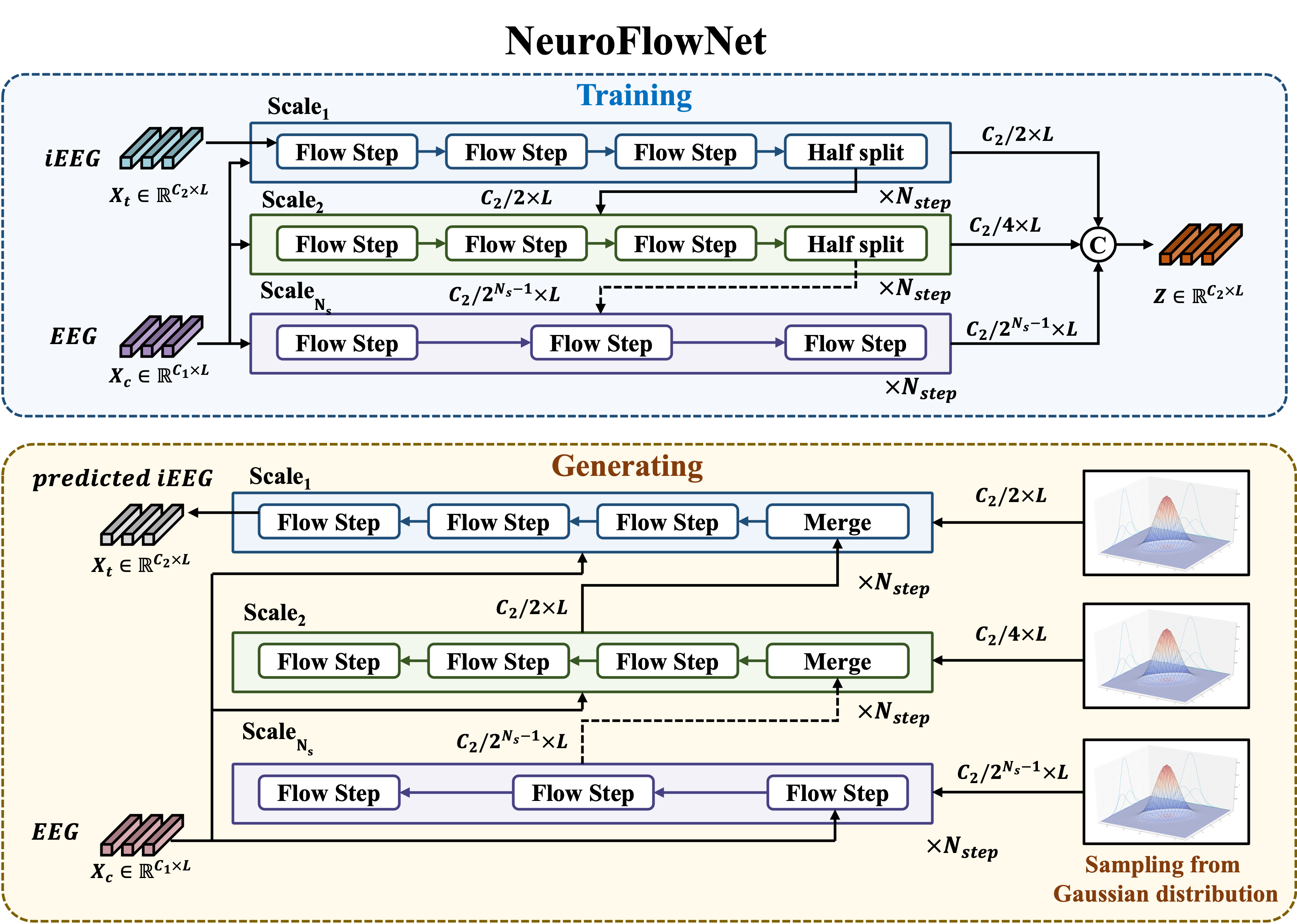}
    \caption{Overview of NeuroFlowNet architecture. The model consists of multiple scales, each containing a series of flow steps that transform the iEEG data conditioned on the EEG data. The final latent variable $Z$ is obtained by concatenating the outputs from all scales.}
    \label{fig:neuroflownet_architecture}
\end{figure*}

The proposed model, NeuroFlowNet, demonstrated in Figure~\ref{fig:neuroflownet_architecture}, is designed to generate iEEG signals from non-invasive EEG signals by learning the cinditional probability distribution of iEEG given EEG, $p(X_t|X_c)$, where $X_t \in \mathbb{R}^{C_{out} \times L}$ represents the target iEEG signals and $X_c \in \mathbb{R}^{C_{in} \times L}$ represents the conditioning EEG signals, with $C_{out}$ and $C_{in}$ being the number of channels of iEEG and EEG, respectively, and $L$ denoting the temporal length of the signals. NeuroFlowNet is architected as a multi-scale conditional normalizing flow, which transforms the complex distribution of iEEG data, conditioned on EEG data, into a simple, tractable base distribution (typically a standard multivariate Gaussian) via a sequence of invertible and differentiable transformations. The conditioning on EEG signals $X_c$ is achieved through a series of coupling layers, where the transformation is conditioned on the EEG data. The model is trained using a maximum likelihood estimation approach, optimizing the parameters of the transformations to maximize the likelihood of the observed iEEG data given the EEG data.

\subsection{Core Invertible Transformation: Conditional Affine Coupling Layer}
The core of the NeuroFlowNet architecture is the conditional affine coupling layer, which allows for the transformation of the iEEG data while conditioning on the EEG data. The coupling layer operates by splitting the input data into two parts: one part is transformed while the other part through an identity transformation. Specifically, given an input vector $x \in \mathbb{R}^{C \times L}$ at any stage of the flow, it is first split along the channel dimension into two halves, $x_a\in\mathbb{R}^{C/2\times L}$ and $x_b\in\mathbb{R}^{C/2\times L}$. The first half, $x_a$, is is passed through an identity transformation, while the second half, $x_b$, is transformed using a conditional affine transformation parameterized by a neural network. The transformation can be expressed as:
\begin{equation}
    z_b = (x_b + \tau) \odot \text{exp}(\sigma)
\end{equation}
where $\odot$ denotes element-wise multiplication. The translation parameters $\tau \in \mathbb{R}^{C/2 \times L}$ and the log-scale parameters $\sigma \in \mathbb{R}^{C/2 \times L}$ are produced by a conditional neural network, denoted as $g_{cond}$, which takes $x_a$ and the conditioning EEG data $X_c$ as inputs, and outputs the parameters $\tau$ and $\sigma$:
\begin{equation}
    \begin{bmatrix}
        \tau \\
        \sigma
    \end{bmatrix} = g_{cond}(x_a, X_c)
\end{equation}

The conditional neural network $g_{cond}$ is designed to capture the complex relationships between the iEEG and EEG data, allowing for a more accurate transformation of the iEEG data based on the conditioning EEG data. The $g_{cond}$ network is composed of an initial couple of convolutional layers with kernel size of 3 and 1, respectively, both followed by a ReLU activation function. Specifically, the first convolutional layers with kernel size of 3 captures local temporal features from the input data and projects the input to a higher-dimensional feature space $\mathbb{R}^{C_h \times L}$, and the second convolutional layer with kernel size of 1 serves as a channel-wise linear transformation, further processing the features while maintaining the temporal resolution. The output of the second convolutional layer is then fed into a multi-head self-attention mechanism to capture long-range dependencies in the data, which is crucial for modeling the complex temporal dynamics of iEEG signals.

The multi-head self-attention mechanism allows the model to capture multiple relationships between the input data and the conditioning data by using multiple sets of query, key, and value matrices, each with different learned parameters. The output of the attention mechanism is then passed through a final convolutional layer with kernel size of 3 to produce the final parameters $\tau$ and $\sigma$. For numerical stability, translation parameters $\tau$ is scaled by a hyperbolic tangent function (tanh) to ensure that the values are bounded between -1 and 1, while the log-scale parameters $\sigma$ are clipped to a range of [-5, 5] to prevent numerical instability during the training process. The log-determinant of the Jacobian for this affine transformation is given by the sum of the log-scale factors:
\begin{equation}
    \log\left|\det\frac{\partial z_b}{\partial x_b}\right|=\sum_{i,j}\sigma_{i,j}
\end{equation}

The output of the coupling layer is $z=\text{concatenate}(x_a,z_b)$. The inverse transformation is obtained by applying the inverse of the affine transformation, which can be expressed as:
\begin{equation}
    x_b = \text{exp}(-\sigma) \odot (z_b - \tau)
\end{equation}
where the inverse transformation is applied to the transformed data $z_b$ to recover the original data $x_b$. 

The inverse transformation is also computed in a similar manner, where the translation and log-scale parameters are obtained from the same conditional neural network $g_{cond}$, but with the input data being the transformed data $z_a$ and the conditioning EEG data $X_c$, i.e.:
\begin{equation}
    \begin{bmatrix}
        \tau \\
        \sigma
    \end{bmatrix} = g_{cond}(z_a, X_c)
\end{equation}

\subsection{Flow Step Composition}
\label{sec:flow_step}

\begin{figure*}
    \centering
    \includegraphics[width=0.8\textwidth]{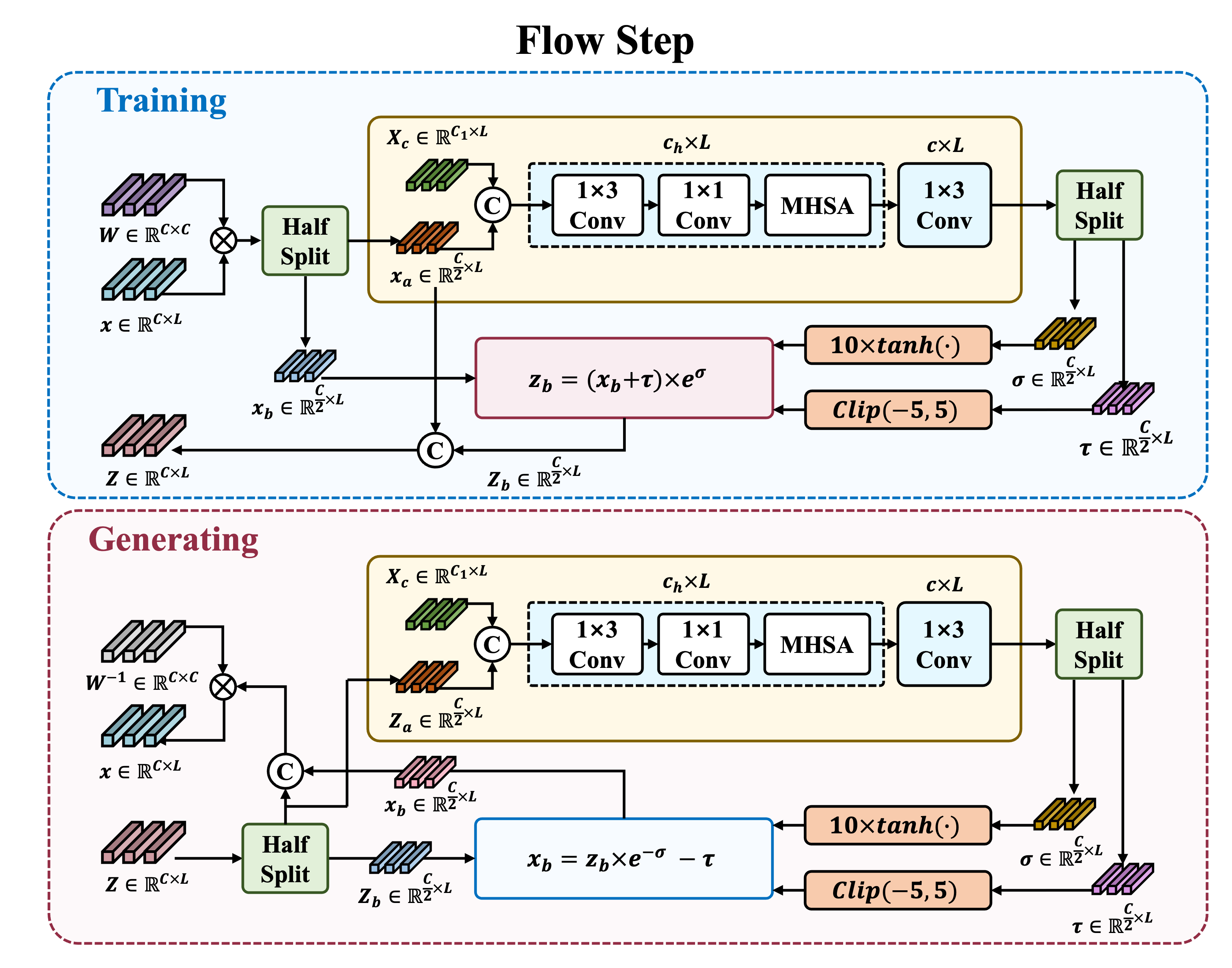}
    \caption{Flow step composition in NeuroFlowNet. Each flow step consists of an invertible $1\times 1$ convolution followed by a conditional affine coupling layer, which transforms the input data while conditioning on the EEG data $X_c$.}
    \label{fig:flow_step}
\end{figure*}

As demonstrated in Figure~\ref{fig:flow_step}, each flow step within NeuroFlowNet combines the conditional affine coupling layer with an invertible channel-mixing operation to enhance the expressive power of the model. Specifically, before the input $x$ is fed into the affine coupling layer, it undergoes an invertible $1\times 1$ convolutional operation, a channel-wise linear transformation, $y=Wx^{\prime}$, where $x^{\prime}$ is the input to the flow step and $W\in\mathbb{R}^{C\times C}$ is a learnable weight matrix, which is initialized as an orthonormal matrix (via QR decomposition of a random Gaussian matrix) to ensure invertibility from the start of training and promote stability. This $1\times 1$ convolution allows for information to be mixed across channels at each temporal location independently. The log-determinant of the Jacobian for this transformation is:
\begin{equation}
    \log\left|\det\frac{\partial y}{\partial x^{\prime}}\right|=L\cdot\log|\det W|
\end{equation}

The total log-determinant for a single flow step is the sum of the log-determinants from the $1\times 1$ convolution and the subsequent conditional affine coupling layer. The inverse of the flow step involves applying the inverse affine coupling transformation followed by multiplication with $W^{-1}$.

\subsection{Multi-Scale Architecture}
The proposed NeuroFlowNet employs a multi-scale architecture to effectively model features and dependencies at varying resolutions within the iEEG signals. The model comprises $N_S$ hierarchical scales. At each scale $k \in \{1, \ldots, N_S\}$, the input, target iEEG signal $X_t$ at the first scale or the output from the preceding finer scale $(k-1)$, is processed through a sequence of $N_{\text{steps}}$ identical flow steps, as detailed in Section~\ref{sec:flow_step}, with all transformations are conditioned on the EEG signal $X_c$.

Following the flow steps at scale $k$, the resulting tensor $y^{(k)} \in \mathbb{R}^{C_k \times L}$ is split along the channel dimension. A subset of channels, $z_{\text{part}}^{(k)} \in \mathbb{R}^{C_k/2 \times L}$, is extracted and designated as part of the final latent variable $Z$. The remaining channels, $x_{\text{in}}^{(k+1)} \in \mathbb{R}^{C_k/2 \times L}$, are propagated as input to the subsequent, typically coarser, scale $(k+1)$. The number of channels $C_k$ is halved at each successive scale, facilitating a progressive abstraction of features.

This hierarchical processing strategy enables the model to decompose the learning task, capturing fine-grained details at the initial scales and progressively more abstract, global representations at deeper levels. The final output from the coarsest scale, $z_{\text{final}}^{(N_S)} \in \mathbb{R}^{C_{N_S} \times L}$, also contributes to the complete latent representation. The full latent variable $Z$ is obtained by concatenating all extracted components and the final output:
\begin{equation}
Z = \text{concatenate}(z_{\text{part}}^{(1)}, z_{\text{part}}^{(2)}, \ldots, z_{\text{part}}^{(N_S - 1)}, z_{\text{final}}^{(N_S)}) \in \mathbb{R}^{C_{out} \times L}
\label{eq:latent_variable}
\end{equation}
where $C_{out} = \sum_{k=1}^{N_S} C_k/2 + C_{N_S}$ is the total number of channels in the latent variable $Z$

\subsection{Training and Inference}
\subsubsection{Training}
NeuroFlowNet is trained by maximizing the exact log-likelihood of the iEEG data $X_t$ given the EEG context $X_c$. Algorithm~\ref{alg:training} outlines the training procedure. Let $f$ denote the overall transformation from the data space to the latent space, such that $Z = f(X_t, X_c)$. The log-likelihood of the iEEG data can be expressed as:
\begin{equation}
    \log p(X_t \mid X_c) = \log p_Z(Z) + \sum_{i=1}^{M} \log \left| \det \frac{\partial z_i}{\partial z_{i-1}} \right|
\end{equation}
where $M$ is the total number of elementary invertible transformations that map the input $X_t$ to the final latent variable $Z$ (as defined in Equation~\ref{eq:latent_variable}), $z_0 = X_t$ and $z_i$ is the output of the $i$-th transformation acting on its input $z_{i-1}$, then the final result of these $M$ transformations is $Z$. The term $\frac{\partial z_i}{\partial z_{i-1}}$ is the Jacobian matrix of this $i$-th transformation. $p_Z(Z)$ is the log-probability density of $Z$ under a standard multivariate Gaussian distribution $\mathcal{N}(0, I)$, which serves as the base distribution. The sum term $\sum_{i=1}^{M} \log \left| \det \frac{\partial z_i}{\partial z_{i-1}} \right|$ is the accumulated sum of the log-determinants of these Jacobians for all constituent invertible transformations ($1\times 1$ convolutions and affine coupling layers) across all flow steps and scales. 

The model parameters are optimized by maximizing this log-likelihood objective:
\begin{equation}
    \theta^* = \arg\max_{\theta} \log p(X_t \mid X_c; \theta)
\end{equation}
where $\theta$ represents the parameters of the model, including the weights of the conditional neural network $g_{cond}$ and the weights of the $1\times 1$ convolutions.

\begin{algorithm}[h]
\caption{Training Phase of NeuroFlowNet (Forward \& Parameter Update)}
\label{alg:training}
\begin{algorithmic}[1]
\Require Batch $\{X_c,X_t\}$, model parameters $\theta$, per-scale splits $\{S_k\}_{k=1}^{N_S}$, time length $L$
\Ensure Updated parameters $\theta$ by maximizing $\log p(X_t\mid X_c)$
\State Initialize $\log\_det\_total \leftarrow \mathbf{0}_{B}$, $o \leftarrow X_t$, $\mathcal{Z}\leftarrow\varnothing$
\For{$k=1 \to N_S$} 
    \State $\log\_det_k \leftarrow \mathbf{0}_{B}$
    \For{$i=1 \to N_{\text{steps}}$} \Comment{FlowStep $i$ in scale $k$}
        \State $u \leftarrow \text{1x1conv}_{W^{(k)}_i}(o)$
        \State $\Delta_{\mathrm{conv}} \leftarrow L\cdot\operatorname{slogdet}\bigl(W^{(k)}_i\bigr)$ \Comment{scalar}
        \State $[u_a,u_b]\leftarrow \text{split\_half}(u)$
        \State $[\tau,\sigma] \leftarrow g_{\text{cond}}(u_a, X_c; \theta_{\text{aff}})$
        \State $\tau\leftarrow 10\cdot\tanh(\tau),\ \sigma\leftarrow\operatorname{clamp}(\sigma,-5,5)$
        \State $z_b \leftarrow (u_b + \tau)\odot \exp(\sigma)$
        \State $o \leftarrow \operatorname{concat}(u_a, z_b)$
        \State $\Delta_{\mathrm{affine}} \leftarrow \sum_{c,t} \sigma_{c,t}$ \Comment{per-sample vector of size $B$}
        \State $\log\_det_k \leftarrow \log\_det_k + (\Delta_{\mathrm{conv}} + \Delta_{\mathrm{affine}})$
    \EndFor
    \State $z^{(k)} \leftarrow o[:,\,1:S_k,\,:]$ \Comment{extract latent part}
    \State $o \leftarrow o[:,\,S_k+1:C_k,\,:]$ \Comment{remaining channels to next scale}
    \State append $z^{(k)}$ to $\mathcal{Z}$
    \State $\log\_det\_total \leftarrow \log\_det\_total + \log\_det_k$
\EndFor
\State append final $o$ to $\mathcal{Z}$ \Comment{final leftover latent}
\State $Z \leftarrow \operatorname{concat}(\mathcal{Z},\ \text{channel})$
\State $\log p_Z \leftarrow -\tfrac{1}{2}\sum_{c,t}\bigl(Z^2 + \log(2\pi)\bigr)$ \Comment{per-sample}
\State $\mathcal{L} \leftarrow -\frac{1}{B}\sum_{b=1}^B\bigl(\log p_Z^{(b)} + \log\_det\_total^{(b)}\bigr)$
\State Backpropagate $\mathcal{L}$ and update $\theta$ (e.g., AdamW step)
\end{algorithmic}
\end{algorithm}

\subsubsection{Inference (iEEG Generation)}
To generate iEEG samples $\hat{X}_t$ conditioned on a given EEG signal $X_c$, the inverse transformation $f^{-1}$ is employed. Algorithm~\ref{alg:inference} outlines the inference procedure. First, a sample $\hat{Z}$ is drawn from the base Gaussian distribution $p_Z(Z)$. This $\hat{Z}$ is then partitioned into segments $\hat{z}^{(1)}_{part}, \ldots, \hat{z}^{(N_S-1)}_{part}, \hat{z}^{(N_S)}_{final}$ corresponding to the multi-scale splitting performed during the forward pass. The generation process starts from the coarsest scale $N_S$. The segment $\hat{z}^{(N_S)}_{final}$ is transformed by applying the inverse of the flow steps within the $N_S$-th scale block. For each subsequent finer scale $k$ (iterating from $N_S - 1$ down to $1$), the corresponding latent segment $\hat{z}^{(k)}_{part}$ is concatenated with the output from the inverse transformation of the preceding coarser scale $(k + 1)$. This combined tensor is then passed through the inverse flow steps of scale $k$. All inverse transformations are conditioned on the EEG context $X_c$. The final output of this sequential inverse process, after passing through the inverse of the first scale block, is the generated iEEG sample $\hat{X}_t = f^{-1}(\hat{Z}; X_c)$.

\begin{algorithm}[t]
\caption{Inference / Sampling Phase of NeuroFlowNet (Inverse Mapping)}
\label{alg:inference}
\begin{algorithmic}[1]
\Require Conditioning $X_c$, model parameters $\theta$, per-scale splits $\{S_k\}_{k=1}^{N_S}$, optional $\varepsilon\in\mathbb{R}^{B\times D\times L}$
\Ensure Generated iEEG samples $\hat{X}_t$
\If{no $\varepsilon$ provided}
    \State sample $\varepsilon\sim\mathcal{N}(0,I)$ \Comment{shape matches total latent channels}
\EndIf
\State split $\varepsilon$ into parts $\{\tilde z^{(1)},\dots,\tilde z^{(N_S)},\tilde z^{(N_S+1)}\}$ according to $\{S_k,\ldots\}$
\State $o \leftarrow \tilde z^{(N_S+1)}$ \Comment{start from coarsest latent}
\For{$k=N_S \to 1$} \Comment{reverse through scales}
    \State $o \leftarrow \operatorname{concat}(\tilde z^{(k)},\, o)$ \Comment{reconstruct full channel tensor at scale $k$}
    \For{each FlowStep $i=N_{\text{steps}},\dots,1$ (reversed order)}
        \State $[z_a,z_b]\leftarrow \text{split\_half}(o)$
        \State $[\tau,\sigma] \leftarrow g_{\text{cond}}(z_a, X_c; \theta_{\text{aff}})$
        \State $\tau\leftarrow 10\cdot\tanh(\tau),\ \sigma\leftarrow\operatorname{clamp}(\sigma,-5,5)$
        \State $u_b \leftarrow \exp(-\sigma)\odot( z_b - \tau )$ \Comment{affine inverse}
        \State $u \leftarrow \operatorname{concat}(z_a, u_b)$
        \State $o \leftarrow \text{1x1conv}_{W^{-1}}(u)$ \Comment{apply inverse 1x1 conv}
    \EndFor
\EndFor
\State \textbf{Output: } $\hat{X}_t \leftarrow o$
\end{algorithmic}
\end{algorithm}

\section{Experiments}
\subsection{Dataset}

\begin{figure*}[h]
    \centering
    \includegraphics[width=0.95\textwidth]{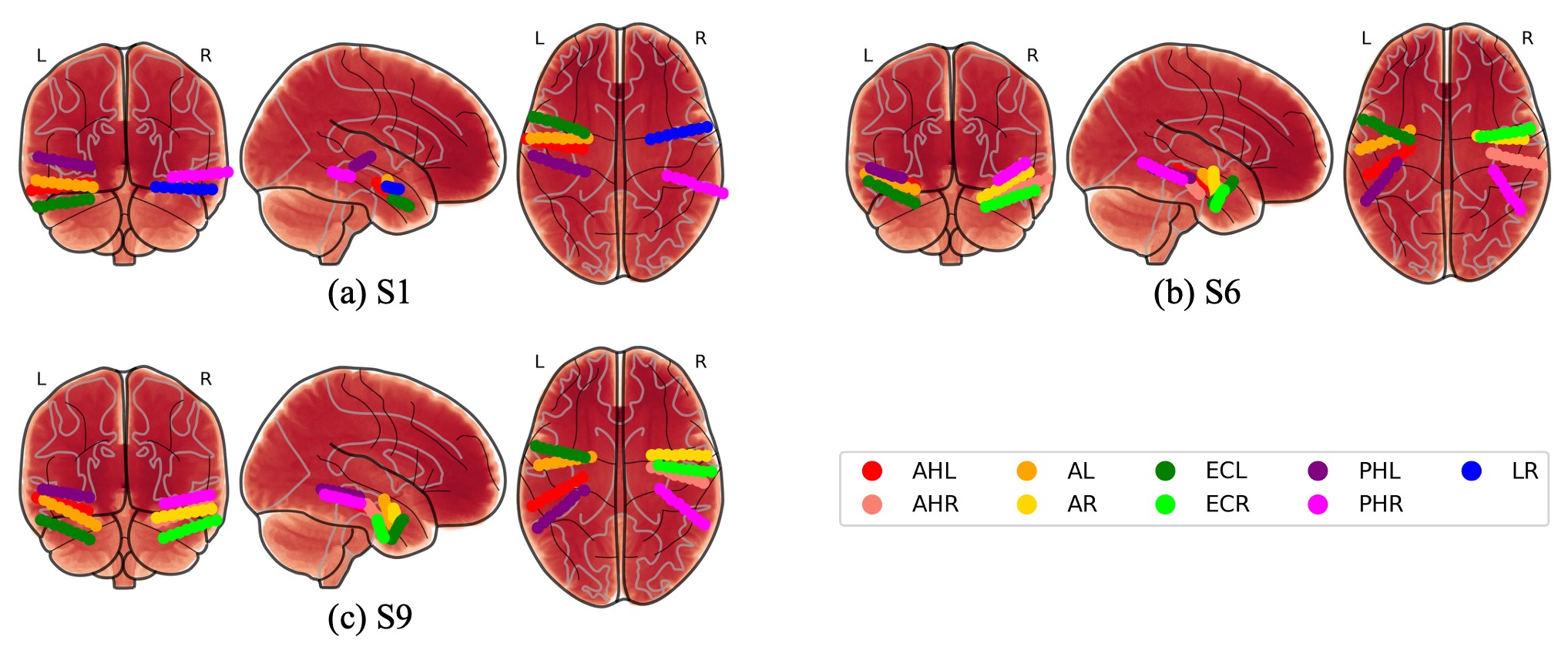}
    \caption{Medial temporal lobe (MTL) depth electrode locations for subjects used in this study (S1, S6 and S9). Each panel shows coronal, sagittal, and axial views of depth electrode trajectories within MTL subregions. Colors indicate anatomical targets: anterior hippocampus (AHL/AHR, red/light red), amygdala (AL/AR, orange/yellow), entorhinal cortex (ECL/ECR, green/light green), parahippocampal gyrus (PHL/PHR, purple/magenta), and lateral rhinal area (LR, blue). Each trajectory consists of eight electrode contacts positioned along a single stereotactic path.}
    \label{fig:MTL_electrode_map}
\end{figure*}

This study utilizes an electrophysiological dataset recorded from nine epilepsy patients during a verbal working memory task, as introduced by Boran et al.~\cite{boran2019persistent}, which is publicly available at \url{https://gin.g-node.org/USZ_NCH/Human_MTL_units_scalp_EEG_and_iEEG_verbal_WM}. The dataset includes simultaneous EEG recorded using the 10-20 system and iEEG obtained via depth microelectrodes implanted in the medial temporal lobe (MTL). The EEG and iEEG signals are originally sampled at 256 Hz and 4 kHz, respectively, and later resampled to 200 Hz and 2 kHz. It also contains waveforms and spike times of 1526 single units, along with MNI coordinates in MNI152 space and anatomical labels for the intracranial electrodes. 

The nine subjects (five female, four male) are all right-handed and aged between 18 and 56. Their pathologies include hippocampal sclerosis, gliosis, xanthoastrocytoma, brain contusion, and focal cortical dysplasia. Each participant has a varying number and location of implanted depth electrodes within the MTL. Due to significant inter-subject variability in electrode implantation (both in terms of anatomical coverage and number of recorded units), only subjects S1, S6, and S9 were included in this study. These three subjects were selected because (i) they have comprehensive electrode coverage across key MTL subregions of interest, (ii) they exhibit a sufficient number of well-isolated single units for reliable statistical analysis, and (iii) their recording sessions contained minimal artifacts and consistent task performance. Due to clinical considerations and individualized implantation plans, the exact anatomical locations of intracranial electrodes differ across subjects~\cite{boran2019persistent}. To illustrate this variability, Figure~\ref{fig:MTL_electrode_map} presents a 3D schematic of depth electrode trajectories in the MTL for three subjects (S1, S6 and S9) used in this study.

Each subject completes between two and seven recording sessions. Table~\ref{tab:subject_characteristics} summarizes the characteristics of the subjects, including their personal information, pathologies, implanted scalp and intracranial electrodes, and number of sessions. The EEG signals are recorded using the international 10-20 system, with electrodes placed at standard locations such as Fp1, Fp2, F7, F3, Cz, Pz, O1, and O2. The iEEG signals are recorded from depth microelectrodes targeting multiple MTL subregions: anterior hippocampus (AHL/AHR), amygdala (AL/AR), entorhinal cortex (ECL/ECR), and parahippocampal gyrus (PHL/PHR), as well as lateral rhinal area (LR). Each trajectory consisted of eight microelectrode contacts arranged along a single stereotactic path.

The dataset provides a rich source of electrophysiological data, allowing for the exploration of cross-modal generation of iEEG signals from EEG signals. The iEEG signals are used as the target signals \( X_t \) for model training and evaluation, while the EEG signals are used as the conditioning input \( X_c \). The dataset is suitable for training and evaluating the NeuroFlowNet model, as it contains a diverse set of subjects with varying pathologies and electrode placements. The model aims to learn the complex relationships between the non-invasive EEG signals and the invasive iEEG signals, enabling the generation of iEEG signals from EEG data.

\begin{table*}[h]
\centering
\caption{Subject Characteristics}
\label{tab:subject_characteristics}
\resizebox{\textwidth}{!}{
\begin{tabular}{cccccc}
\toprule
\textbf{No.} & \textbf{Age} & \textbf{Sex} & \textbf{Pathology} & \textbf{Implanted electrodes} & \textbf{Sessions} \\
\midrule
\multirow{2}{*}{1} & \multirow{2}{*}{24} & \multirow{2}{*}{F} & \multirow{2}{*}{Xanthoastrocytoma WHO II} & F3, F4, C3, C4, P3, P4, O1, O2, F7, F8, T3, T4, T5, T6, Fz, Cz, Pz, A1, A2 & \multirow{2}{*}{4} \\
    & & & & AHL, AL, ECL, LR, PHL, PHR & \\
\multirow{2}{*}{2} & \multirow{2}{*}{39} & \multirow{2}{*}{M} & \multirow{2}{*}{Gliosis} & F3, F4, C3, C4, O1, O2, A1, A2 & \multirow{2}{*}{7} \\
    & & & & AHL, AHR, AL, AR, ECL, ECR, PHL, PHR & \\
\multirow{2}{*}{3} & \multirow{2}{*}{18} & \multirow{2}{*}{F} & \multirow{2}{*}{Hippocampal sclerosis} & F3, F4, C3, C4, O1, O2, A1, A2 & \multirow{2}{*}{3} \\
    & & & & AHL, AHR, AL, ECL, PHL & \\
\multirow{2}{*}{4} & \multirow{2}{*}{28} & \multirow{2}{*}{M} & \multirow{2}{*}{Brain contusion} & F3, F4, C3, C4, P3, P4, O1, O2, F7, F8, T3, T4, T5, T6, Fz, Cz, Pz, A1, A2 & \multirow{2}{*}{2} \\
    & & & & AHL, AHR, AL, AR, ECL, ECR, PHL, PHR & \\
\multirow{2}{*}{5} & \multirow{2}{*}{20} & \multirow{2}{*}{F} & \multirow{2}{*}{Focal cortical dysplasia} & Fp1, Fp2, F3, F4, C3, C4, P3, P4, O1, O2, F7, F8, T4, T5, T6, Fz, Cz, Pz, A1, A2 & \multirow{2}{*}{3} \\
    & & & & AHL, AL, DRR, PHR & \\
\multirow{2}{*}{6} & \multirow{2}{*}{31} & \multirow{2}{*}{M} & \multirow{2}{*}{Hippocampal sclerosis} & Fp1, Fp2, F3, F4, C3, C4, O1, O2, A1, A2 & \multirow{2}{*}{7} \\
    & & & & AHL, AHR, AL, AR, ECL, ECR, PHL, PHR & \\
\multirow{2}{*}{7} & \multirow{2}{*}{47} & \multirow{2}{*}{M} & \multirow{2}{*}{Hippocampal sclerosis} & F3, F4, C3, C4, O1, O2, A1, A2 & \multirow{2}{*}{4} \\
    & & & & AHL, AHR, AL, AR, ECL, ECR, PHL, PHR & \\
\multirow{2}{*}{8} & \multirow{2}{*}{56} & \multirow{2}{*}{F} & \multirow{2}{*}{Hippocampal sclerosis} & F3, F4, C3, C4, P3, P4, O1, O2, F7, F8, T3, T4, T5, T6, Fz, Cz, Pz, A1, A2 & \multirow{2}{*}{5} \\
    & & & & AHL, AHR, AL, AR, ECL, ECR, PHL, PHR & \\
\multirow{2}{*}{9} & \multirow{2}{*}{19} & \multirow{2}{*}{F} & \multirow{2}{*}{Hippocampal sclerosis} & F3, F4, C3, C4, O1, O2, A1, A2 & \multirow{2}{*}{2} \\
    & & & & AHL, AHR, AL, AR, ECL, ECR, PHL, PHR & \\
\bottomrule
\end{tabular}
}
\\[1ex]
\footnotesize{
    $^1$ Each iEEG electrode label corresponds to eight depth microelectrode contacts positioned along a single trajectory.\\
    $^2$ Abbreviations for MTL Regions: AHL/AHR - Anterior Hippocampus (Left/Right), AL/AR - Amygdala (Left/Right), ECL/ECR - Entorhinal Cortex (Left/Right), PHL/PHR - Parahippocampal Gyrus (Left/Right), LR - Lateral Rhinal Area, DRR - Dorsal Rhinal Region.
}
\end{table*}

\subsection{Data Preprocessing and Division}
The iEEG and EEG signals are preprocessed to remove artifacts and noise, ensuring high-quality data for model training. The preprocessing steps include wavelet denoising, downsampling, data division, and normalization.

The wavelet denoising is performed using the discrete wavelet transform (DWT) with Symlet wavelets of order 4 (sym4), which helps suppress broadband noise while retaining the main morphological characteristics of the signals. A three-level decomposition is applied to each channel independently, producing the coefficient set \({A_3, D_3, D_2, D_1}\), where \(A_3\) denotes the level-3 approximation coefficients and \(D_\ell\) denotes the detail coefficients at level \(\ell\). The approximation coefficients \(A_3\) are retained without modification, and soft-thresholding is applied to all detail sub-bands \({D_1, D_2, D_3}\). The noise level is estimated using the median absolute deviation (MAD) computed over the concatenation of detail coefficients across levels, \(\hat{\sigma}=\mathrm{median}(|\mathrm{concat}(D_1,D_2,D_3)|)/0.6745\). A universal threshold is then determined based on the segment length (n) (number of time samples) as \(T=\hat{\sigma}\sqrt{2\ln(n)}\), and each detail coefficient (w) is shrunk via soft thresholding \(\mathcal{S}_T(w)=\mathrm{sign}(w)\max(|w|-T,0)\). Finally, the denoised signal is reconstructed by inverse DWT (IDWT) with symmetric boundary extension, and the reconstructed sequence is cropped to the original length to ensure temporal alignment with the original signal. We note that the effect of this denoising procedure on transient waveform morphology was not directly evaluated in the present study. Therefore, the subsequent waveform and spectral evaluations should be interpreted as reconstruction performance relative to the denoised and band-limited target signals, rather than as evidence that all transient morphological features in the original broadband iEEG recordings were preserved.

In this work, to align the temporal grid between modalities for supervised cross-modal learning, we further downsampled iEEG to 200~Hz using an explicit anti-aliasing procedure, which constrains the analyzable bandwidth to $\leq$100~Hz (and our evaluations to 0.5-50~Hz / 8-13~Hz). Specifically, to prevent aliasing, we apply an explicit anti-aliasing low-pass filter prior to decimation: iEEG is first low-pass filtered with a cutoff below the target Nyquist frequency (100~Hz) and a transition band (implemented in a zero-phase manner via forward-backward filtering), and then resampled to 200~Hz using a polyphase/decimation routine. As a consequence, the maximum representable frequency after downsampling is 100~Hz, and all frequency-domain evaluations and claims in this paper are restricted to the preserved low-/mid-frequency range (PSD: 0.5-50~Hz; alpha: 8-13~Hz; see Tables~\ref{tab:performance_results_trial} and~\ref{tab:performance_results_trial_session}).

Each subject participates in multiple recording sessions, and each session consists of multiple trials of the verbal working memory task. Due to the individualized clinical implantation schemes (i.e., subject-specific electrode locations, trajectories, and the number of available contacts/units), the intracranial targets and coverage are not directly comparable across subjects; therefore, a controlled cross-subject generalization setting is not feasible with this dataset. Accordingly, we adopt subject-specific (within-subject) evaluation protocols, where models are trained and validated using data from the same subject. To assess robustness at different granularities, we report two within-subject settings. In the cross-trial setting, within each session, 90\% of trials are randomly assigned to the training set and the remaining 10\% to the evaluation set, so the model is tested on unseen trials from sessions observed during training. In the cross-session setting, entire recording sessions are held out from model fitting and used only for evaluation, so the model is tested on unseen sessions from the same subject. The cross-trial and cross-session results are reported in Tables~\ref{tab:performance_results_trial} and~\ref{tab:performance_results_trial_session}, respectively. After splitting, the recordings are further divided into smaller segments of 200 samples (1 second) using a sliding-window approach with a 90\% overlap for the training set, while the evaluation set is divided into segments of 200 samples with no overlap. Z-score normalization is applied to each segment, ensuring that the mean and standard deviation of the signals are zero and one, respectively. This normalization step is crucial for stabilizing the training process and improving the model's performance.

\subsubsection{Regression baseline comparison setup}
\label{sec:connectivity_baseline_setup}

To evaluate whether NeuroFlowNet preserves inter-channel correlation structure more effectively than deterministic alternatives, we additionally compare it with four aligned sequence-to-sequence regression baselines: (i) a pointwise linear mapping implemented as a $1\times1$ convolution (Linear), (ii) a shallow 1D CNN regressor (Shallow CNN), (iii) a non-causal 1D U-Net regressor (1D U-Net), and (iv) a tiny Transformer regressor (Tiny Transformer). All models are trained and evaluated using the same subject-specific setting (S1, S6, S9), the same trial-level split (90\%/10\%), and the same segmentation strategy (1-second windows of 200 samples; training windows with overlap and validation windows without overlap). This ensures that any performance differences are attributable to modeling choices rather than data partitioning or preprocessing.

For each 1-second segment, we compute the Pearson inter-channel correlation matrix $\mathbf{R}\in\mathbb{R}^{C\times C}$ from the iEEG channels over the time axis. Let $\mathbf{R}^{\text{pred}}$ and $\mathbf{R}^{\text{true}}$ denote the predicted and ground-truth correlation matrices, respectively. We report two matrix-level errors:
\begin{equation}
\mathcal{E}_{\mathrm{MAE}}(\mathbf{R}) = \frac{1}{C^2}\left\lVert \mathbf{R}^{\text{pred}}-\mathbf{R}^{\text{true}}\right\rVert_{1},\quad
\mathcal{E}_{F}(\mathbf{R})=\left\lVert \mathbf{R}^{\text{pred}}-\mathbf{R}^{\text{true}}\right\rVert_{F},
\end{equation}
where $\lVert\cdot\rVert_{1}$ is the element-wise $\ell_1$ norm and $\lVert\cdot\rVert_F$ is the Frobenius norm. Metrics are first averaged within each subject and then across subjects (S1, S6, S9), reported as mean$\pm$std. Lower values indicate better preservation of inter-channel correlation structure in the reconstructed band-limited signals.

\subsection{Hyperparameter Selection and Ablation Study}
\label{sec:hyperparameters}
\begin{figure*}
    \centering
    \begin{subfigure}[b]{0.33\textwidth}
        \centering
        \includegraphics[width=\linewidth]{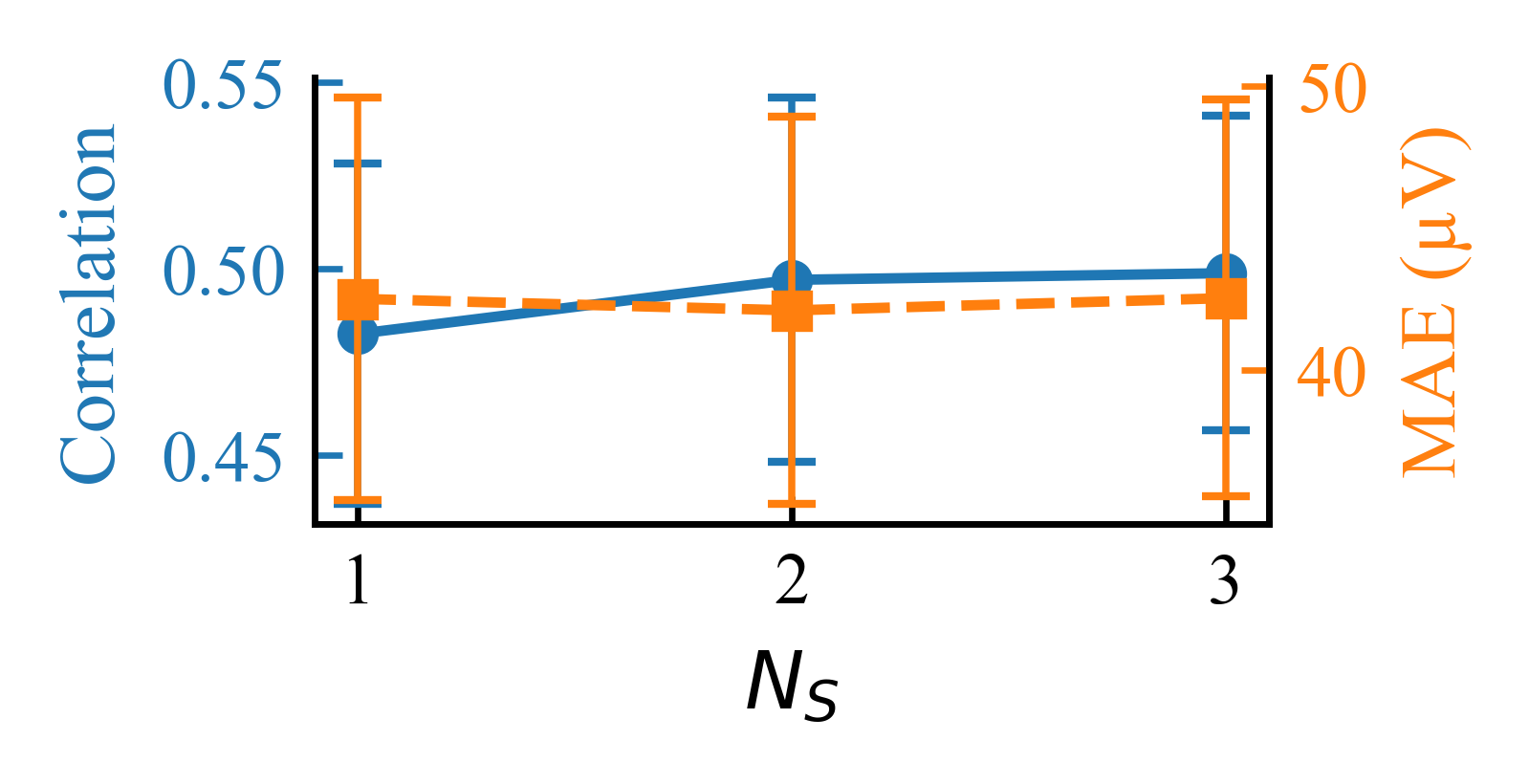}
        \caption{$N_S$ (number of scales)}
    \end{subfigure}
    \begin{subfigure}[b]{0.33\textwidth}
        \centering
        \includegraphics[width=\linewidth]{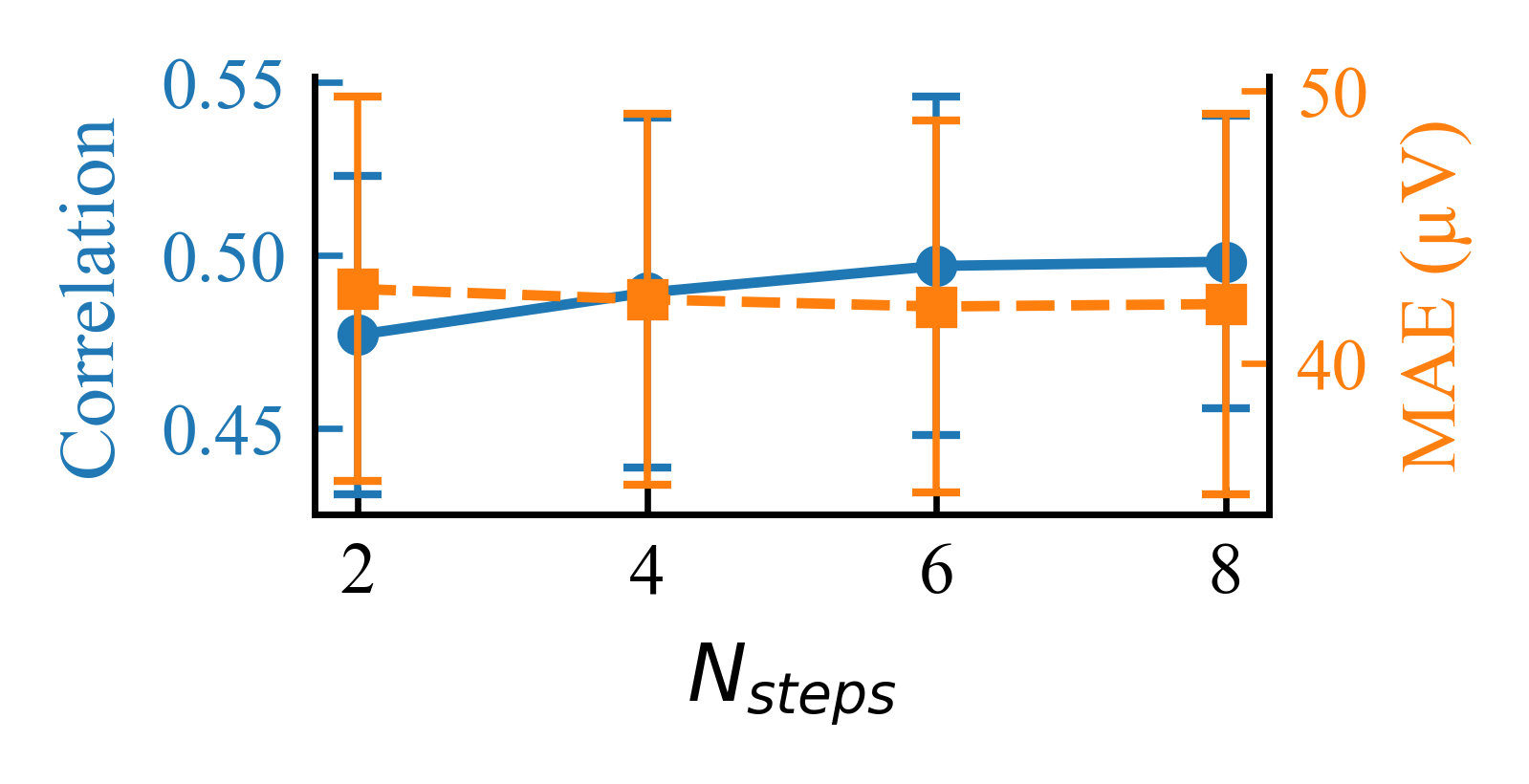}
        \caption{$N_{\text{steps}}$ (flow steps per scale)}
    \end{subfigure}
    \begin{subfigure}[b]{0.33\textwidth}
        \centering
        \includegraphics[width=\linewidth]{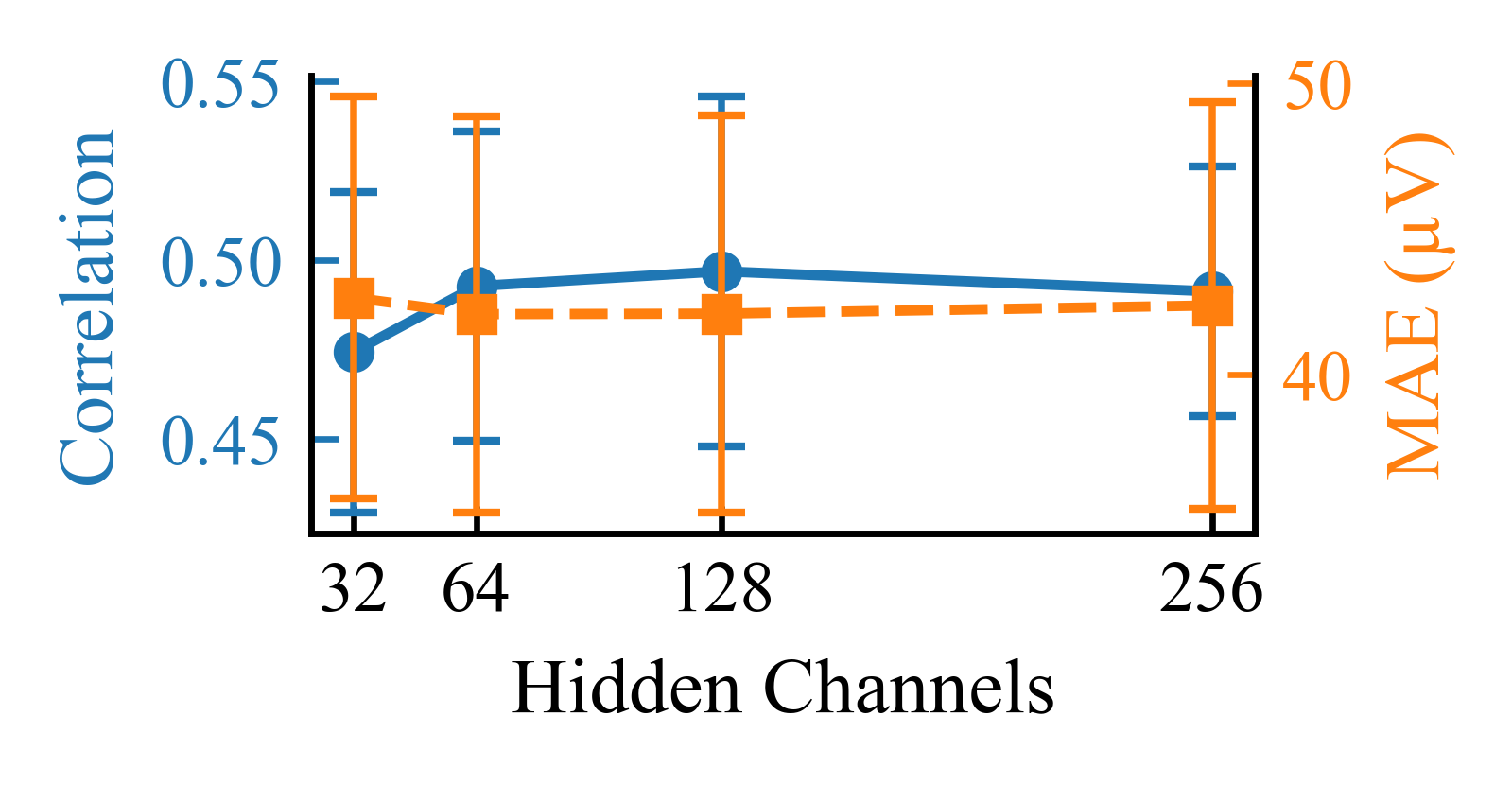}
        \caption{$C_h$ (hidden channels)}
    \end{subfigure}

    \caption{Hyperparameter selection results for NeuroFlowNet. The performance of the model is evaluated on the validation set and averaged across all subjects and electrodes. The hyperparameters include $N_S$, the number of scales (a); $N_{\text{steps}}$, the number of flow steps per scale (b); and $C_h$, the number of channels in the hidden layers of the conditional neural network (c). The results indicate that the model achieves optimal performance with $N_S=2$, $N_{\text{steps}}=6$, and $C_h=128$.}
    \label{fig:hyperparameters}
\end{figure*}

The performance of the NeuroFlowNet model is influenced by several key hyperparameters, including the number of scales $N_S$, the number of flow steps per scale $N_{\text{steps}}$, and the number of channels in the hidden layers of the conditional neural network $C_h$. To determine the optimal values for these hyperparameters, we conducted a systematic evaluation using a one-factor-at-a-time (OFAT) approach. This method involves varying one hyperparameter while keeping the others fixed at their default values, allowing us to isolate the effect of each hyperparameter on model performance. The candidate values for each hyperparameter were selected based on preliminary experiments and domain knowledge. The performance of the model was evaluated on the validation set and averaged across all subjects to ensure that the selected hyperparameters generalize well across different individuals. Specifically, as shown in Figure~\ref{fig:hyperparameters}(a), varying the number of scales $N_S \in \{1,2,3\}$ leads to a small improvement when increasing from $N_S=1$ to $N_S=2$ (higher correlation with lower MAE). Further increasing to $N_S=3$ brings no consistent gain, with performance remaining essentially unchanged within the error bars. As illustrated in Figure~\ref{fig:hyperparameters}(b), we evaluated the number of flow steps per scale $N_{\text{steps}} \in \{2,4,6,8\}$. The correlation improves as $N_{\text{steps}}$ increases up to 6, while MAE shows only minor variation; increasing to $N_{\text{steps}}=8$ yields no clear additional benefit. Finally, as shown in Figure~\ref{fig:hyperparameters}(c), we assessed the effect of the hidden channels $C_h \in \{32,64,128,256\}$ in the conditional network. The model achieves the best (or near-best) validation performance at $C_h=128$, whereas smaller widths underperform slightly and a larger width ($C_h=256$) does not improve results. Based on these observations and considering computational efficiency, we select $N_S=2$, $N_{\text{steps}}=6$, and $C_h=128$ as the hyperparameter setting for training NeuroFlowNet.

\begin{figure}
    \centering
    \includegraphics[width=0.6\textwidth]{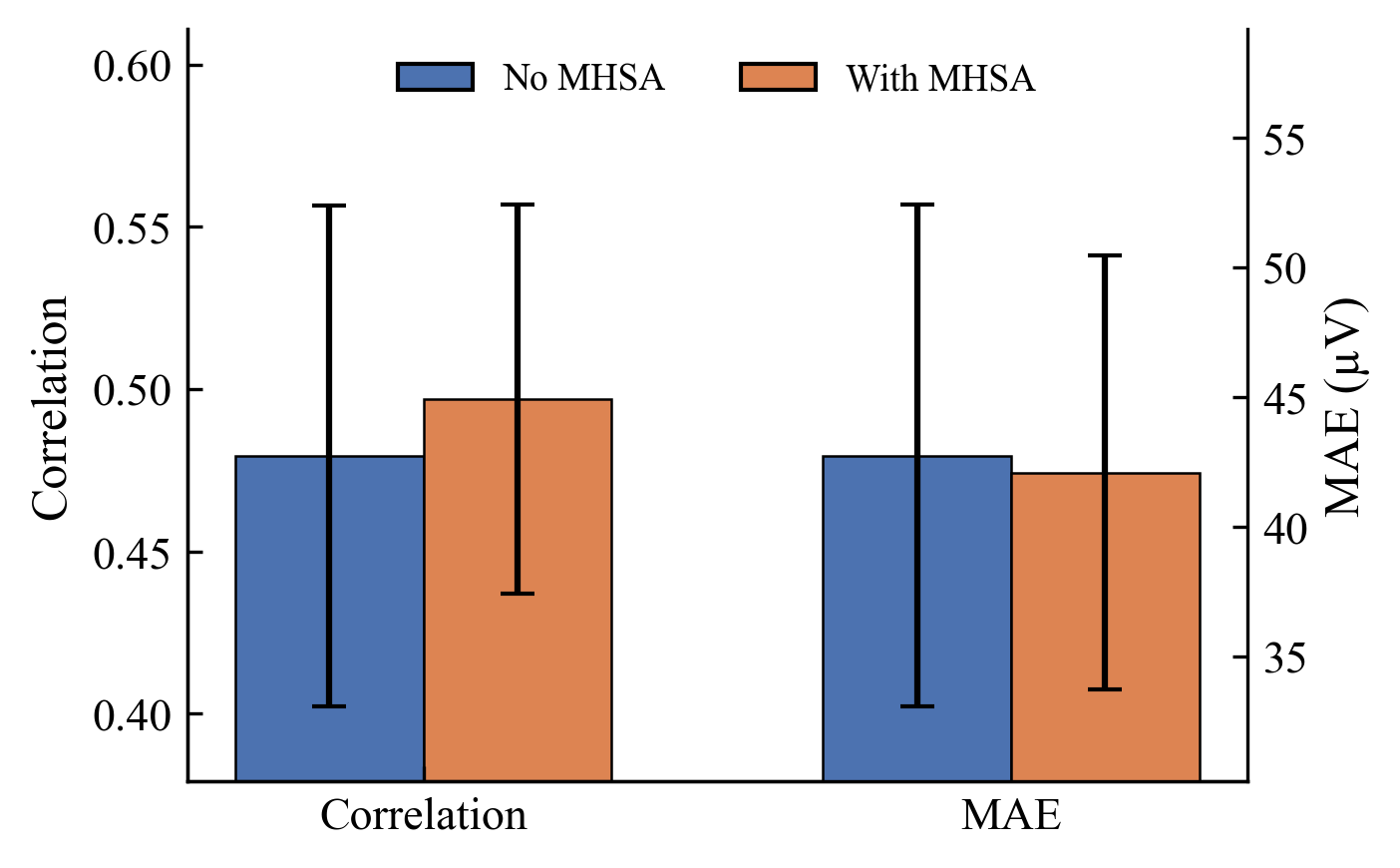}
    \caption{Ablation of the MHSA module in the conditional network $g_{cond}$. Correlation (higher is better) and MAE (lower is better) are reported on the validation set. Bars indicate the mean across subjects and electrodes, and error bars denote the standard deviation across subjects}
    \label{fig:ablation_MHSA}
\end{figure}

To validate the architectural choice of the conditional network $g_{cond}$ in the affine coupling layers, we conduct a controlled ablation by removing the MHSA module from $g_{cond}$ while keeping all other components and training/evaluation settings unchanged. Figure~\ref{fig:ablation_MHSA} summarizes the results on the validation set. Overall, incorporating MHSA yields better performance when averaged across subjects and electrodes. The mean correlation increases from $0.479 \pm 0.077$ (No MHSA) to $0.497 \pm 0.060$ (With MHSA), and the mean MAE decreases from $42.74 \pm 9.66~\mu$V to $42.09 \pm 8.35~\mu$V (mean$\pm$std across subjects and electrodes). These results suggest that MHSA strengthens $g_{cond}$ in capturing cross-modal dependencies between EEG and iEEG, improving reconstruction consistency (higher correlation) and slightly reducing amplitude error (lower MAE) on average. This ablation provides component-level evidence that the attention mechanism is a beneficial design element in $g_{cond}$ for cross-modal generation.

\subsection{Implementation Details}
The NeuroFlowNet model is implemented using PyTorch, a popular deep learning framework. The model architecture consists of multiple scales, each containing a series of flow steps that transform the iEEG data conditioned on the EEG data. The model is trained using the AdamW optimizer with a learning rate of $1\times 10^{-3}$ and a weight decay of $1\times 10^{-2}$. The model is trained on a single NVIDIA RTX 4080 GPU with 16 GB of memory. The batch size is set to 128, and the model is trained for 100 epochs. The $N_S$ (number of scales), $N_{\text{steps}}$ (number of flow steps per scale) and $C_h$ (number of channels in the hidden layers of the conditional neural network) are treated as hyperparameters. Candidate values are chosen empirically and then evaluated on the validation set using OFAT procedure, where one hyperparameter is varied while the others are fixed, as illustrated in Section~\ref{sec:hyperparameters}. The final model architecture is set to have $N_S=2$ scales, with $N_{\text{steps}}=6$ flow steps per scale, and $C_h=128$ channels in the hidden layers of the conditional neural network. The model is trained separately for each subject, and the best-performing model on the validation set is selected. Early stopping is implemented based on the performance on the validation set, with a patience of 10 epochs.

\section{Results and Discussion}

\begin{table*}[h]
\centering
\caption{Cross-trial performance of NeuroFlowNet on iEEG signal generation.}
\label{tab:performance_results_trial}
\resizebox{\textwidth}{!}{
\begin{tabular}{clcccccccc}
\toprule
\multirow{2}{*}{\textbf{Subject}} & \multirow{2}{*}{\textbf{Region}} & \multicolumn{3}{c}{\textbf{Temporal waveform}} & \multicolumn{2}{c}{\textbf{PSD waveform (0.5-50Hz)}} & \multicolumn{3}{c}{\textbf{Alpha PSD waveform (8-13Hz)}} \\
\cmidrule(lr){3-5} \cmidrule(lr){6-7} \cmidrule(lr){8-10}
    & & \textbf{MAE (µV)} & \textbf{Corr} & \textbf{Cosine} & \textbf{Corr} & \textbf{Cosine} & \textbf{Corr} & \textbf{Cosine} & \textbf{Power MAPE (\%)} \\
\midrule\multirow{6}{*}{S1} 
& AHL & 40.66 $\pm$ 21.53 & 0.35 $\pm$ 0.28 & 0.47 $\pm$ 0.24 & 0.66 $\pm$ 0.22 & 0.70 $\pm$ 0.19 & 0.25 $\pm$ 0.53 & 0.70 $\pm$ 0.21 & 2.65 $\pm$ 4.40 \\
& AL & 42.82 $\pm$ 28.98 & 0.39 $\pm$ 0.30 & 0.50 $\pm$ 0.26 & 0.67 $\pm$ 0.23 & 0.71 $\pm$ 0.20 & 0.29 $\pm$ 0.52 & 0.71 $\pm$ 0.21 & 3.33 $\pm$ 7.35 \\
& ECL & 49.15 $\pm$ 27.35 & 0.40 $\pm$ 0.25 & 0.50 $\pm$ 0.21 & 0.66 $\pm$ 0.24 & 0.70 $\pm$ 0.21 & 0.32 $\pm$ 0.52 & 0.72 $\pm$ 0.21 & 6.49 $\pm$ 16.65 \\
& LR & 33.86 $\pm$ 20.08 & 0.60 $\pm$ 0.28 & 0.65 $\pm$ 0.25 & 0.76 $\pm$ 0.23 & 0.79 $\pm$ 0.20 & 0.45 $\pm$ 0.48 & 0.78 $\pm$ 0.18 & 1.24 $\pm$ 2.18 \\
& PHL & 35.68 $\pm$ 15.08 & 0.56 $\pm$ 0.33 & 0.62 $\pm$ 0.28 & 0.74 $\pm$ 0.23 & 0.77 $\pm$ 0.20 & 0.44 $\pm$ 0.49 & 0.77 $\pm$ 0.20 & 1.23 $\pm$ 2.87 \\
& PHR & 82.42 $\pm$ 72.44 & 0.24 $\pm$ 0.29 & 0.39 $\pm$ 0.27 & 0.53 $\pm$ 0.25 & 0.59 $\pm$ 0.22 & 0.23 $\pm$ 0.53 & 0.70 $\pm$ 0.20 & 2.85 $\pm$ 5.36 \\
\midrule\multirow{8}{*}{S6} 
& AHL & 44.08 $\pm$ 27.47 & 0.44 $\pm$ 0.23 & 0.51 $\pm$ 0.21 & 0.61 $\pm$ 0.21 & 0.67 $\pm$ 0.18 & 0.33 $\pm$ 0.51 & 0.74 $\pm$ 0.19 & 1.58 $\pm$ 3.63 \\
& AHR & 40.80 $\pm$ 21.73 & 0.58 $\pm$ 0.21 & 0.63 $\pm$ 0.19 & 0.68 $\pm$ 0.20 & 0.73 $\pm$ 0.17 & 0.48 $\pm$ 0.50 & 0.80 $\pm$ 0.19 & 0.95 $\pm$ 1.88 \\
& AL & 45.78 $\pm$ 19.99 & 0.46 $\pm$ 0.23 & 0.52 $\pm$ 0.21 & 0.63 $\pm$ 0.21 & 0.67 $\pm$ 0.18 & 0.39 $\pm$ 0.48 & 0.76 $\pm$ 0.19 & 1.75 $\pm$ 2.97 \\
& AR & 26.77 $\pm$ 9.21 & 0.74 $\pm$ 0.16 & 0.77 $\pm$ 0.14 & 0.78 $\pm$ 0.17 & 0.81 $\pm$ 0.15 & 0.67 $\pm$ 0.38 & 0.87 $\pm$ 0.14 & 0.61 $\pm$ 1.01 \\
& ECL & 53.71 $\pm$ 21.22 & 0.49 $\pm$ 0.23 & 0.54 $\pm$ 0.21 & 0.67 $\pm$ 0.21 & 0.71 $\pm$ 0.18 & 0.37 $\pm$ 0.51 & 0.75 $\pm$ 0.19 & 1.75 $\pm$ 3.41 \\
& ECR & 34.92 $\pm$ 14.32 & 0.64 $\pm$ 0.17 & 0.68 $\pm$ 0.16 & 0.71 $\pm$ 0.20 & 0.75 $\pm$ 0.18 & 0.59 $\pm$ 0.43 & 0.84 $\pm$ 0.15 & 1.05 $\pm$ 2.48 \\
& PHL & 25.73 $\pm$ 15.79 & 0.39 $\pm$ 0.29 & 0.49 $\pm$ 0.25 & 0.60 $\pm$ 0.21 & 0.66 $\pm$ 0.18 & 0.32 $\pm$ 0.51 & 0.74 $\pm$ 0.20 & 1.38 $\pm$ 2.51 \\
& PHR & 36.89 $\pm$ 16.07 & 0.60 $\pm$ 0.20 & 0.64 $\pm$ 0.18 & 0.70 $\pm$ 0.19 & 0.74 $\pm$ 0.17 & 0.51 $\pm$ 0.45 & 0.79 $\pm$ 0.18 & 0.84 $\pm$ 1.53 \\
\midrule\multirow{8}{*}{S9} 
& AHL & 29.92 $\pm$ 10.79 & 0.67 $\pm$ 0.19 & 0.73 $\pm$ 0.15 & 0.81 $\pm$ 0.17 & 0.82 $\pm$ 0.16 & 0.36 $\pm$ 0.49 & 0.73 $\pm$ 0.20 & 0.88 $\pm$ 1.24 \\
& AHR & 43.49 $\pm$ 26.44 & 0.38 $\pm$ 0.28 & 0.50 $\pm$ 0.24 & 0.69 $\pm$ 0.23 & 0.71 $\pm$ 0.21 & 0.20 $\pm$ 0.55 & 0.67 $\pm$ 0.22 & 7.20 $\pm$ 14.62 \\
& AL & 35.30 $\pm$ 11.46 & 0.61 $\pm$ 0.19 & 0.66 $\pm$ 0.16 & 0.78 $\pm$ 0.19 & 0.80 $\pm$ 0.17 & 0.34 $\pm$ 0.50 & 0.72 $\pm$ 0.21 & 1.45 $\pm$ 3.01 \\
& AR & 29.48 $\pm$ 21.12 & 0.34 $\pm$ 0.34 & 0.46 $\pm$ 0.30 & 0.64 $\pm$ 0.25 & 0.68 $\pm$ 0.22 & 0.22 $\pm$ 0.54 & 0.67 $\pm$ 0.23 & 4.82 $\pm$ 8.81 \\
& ECL & 38.73 $\pm$ 17.66 & 0.55 $\pm$ 0.25 & 0.61 $\pm$ 0.21 & 0.75 $\pm$ 0.22 & 0.77 $\pm$ 0.20 & 0.34 $\pm$ 0.52 & 0.72 $\pm$ 0.22 & 1.62 $\pm$ 2.73 \\
& ECR & 43.11 $\pm$ 18.69 & 0.44 $\pm$ 0.26 & 0.51 $\pm$ 0.23 & 0.71 $\pm$ 0.22 & 0.74 $\pm$ 0.20 & 0.24 $\pm$ 0.52 & 0.70 $\pm$ 0.21 & 3.46 $\pm$ 5.78 \\
& PHL & 26.60 $\pm$ 14.10 & 0.73 $\pm$ 0.20 & 0.78 $\pm$ 0.17 & 0.84 $\pm$ 0.16 & 0.85 $\pm$ 0.15 & 0.46 $\pm$ 0.46 & 0.77 $\pm$ 0.18 & 0.93 $\pm$ 1.39 \\
& PHR & 38.48 $\pm$ 17.46 & 0.49 $\pm$ 0.26 & 0.55 $\pm$ 0.25 & 0.74 $\pm$ 0.22 & 0.76 $\pm$ 0.20 & 0.33 $\pm$ 0.51 & 0.71 $\pm$ 0.21 & 5.35 $\pm$ 10.49 \\
\bottomrule
\end{tabular}
}
\\[1ex]
\footnotesize{
    $^1$ The results of each region are averaged across all electrodes within that region.\\
    $^2$ Abbreviations for MTL Regions: AHL/AHR - Anterior Hippocampus (Left/Right), AL/AR - Amygdala (Left/Right), ECL/ECR - Entorhinal Cortex (Left/Right), PHL/PHR - Parahippocampal Gyrus (Left/Right), LR - Lateral Rhinal Area.
}
\end{table*}

\begin{table*}[h]
\centering
\caption{Cross-session performance of NeuroFlowNet on iEEG signal generation.}
\label{tab:performance_results_trial_session}
\resizebox{\textwidth}{!}{
\begin{tabular}{clcccccccc}
\toprule
\multirow{2}{*}{\textbf{Subject}} & \multirow{2}{*}{\textbf{Region}} & \multicolumn{3}{c}{\textbf{Temporal waveform}} & \multicolumn{2}{c}{\textbf{PSD waveform (0.5-50Hz)}} & \multicolumn{3}{c}{\textbf{Alpha PSD waveform (8-13Hz)}} \\
\cmidrule(lr){3-5} \cmidrule(lr){6-7} \cmidrule(lr){8-10}
    & & \textbf{MAE (µV)} & \textbf{Corr} & \textbf{Cosine} & \textbf{Corr} & \textbf{Cosine} & \textbf{Corr} & \textbf{Cosine} & \textbf{Power MAPE (\%)} \\
\midrule\multirow{6}{*}{S1} 
& AHL & 52.48 $\pm$ 25.26 & 0.32 $\pm$ 0.27 & 0.43 $\pm$ 0.24 & 0.71 $\pm$ 0.22 & 0.74 $\pm$ 0.20 & 0.17 $\pm$ 0.50 & 0.69 $\pm$ 0.19 & 3.00 $\pm$ 6.60 \\
& AL & 67.56 $\pm$ 44.54 & 0.56 $\pm$ 0.23 & 0.63 $\pm$ 0.19 & 0.81 $\pm$ 0.19 & 0.83 $\pm$ 0.17 & 0.21 $\pm$ 0.52 & 0.69 $\pm$ 0.20 & 5.43 $\pm$ 16.05 \\
& ECL & 67.11 $\pm$ 32.60 & 0.46 $\pm$ 0.24 & 0.55 $\pm$ 0.19 & 0.73 $\pm$ 0.23 & 0.76 $\pm$ 0.21 & 0.18 $\pm$ 0.50 & 0.68 $\pm$ 0.20 & 10.32 $\pm$ 21.51 \\
& LR & 46.66 $\pm$ 23.64 & 0.83 $\pm$ 0.16 & 0.84 $\pm$ 0.14 & 0.93 $\pm$ 0.12 & 0.94 $\pm$ 0.11 & 0.40 $\pm$ 0.49 & 0.78 $\pm$ 0.18 & 1.15 $\pm$ 2.07 \\
& PHL & 38.74 $\pm$ 17.30 & 0.84 $\pm$ 0.15 & 0.84 $\pm$ 0.13 & 0.93 $\pm$ 0.12 & 0.93 $\pm$ 0.11 & 0.41 $\pm$ 0.50 & 0.77 $\pm$ 0.19 & 0.71 $\pm$ 1.00 \\
& PHR & 148.69 $\pm$ 70.33 & 0.05 $\pm$ 0.21 & 0.29 $\pm$ 0.27 & 0.46 $\pm$ 0.23 & 0.53 $\pm$ 0.20 & 0.08 $\pm$ 0.52 & 0.67 $\pm$ 0.20 & 5.69 $\pm$ 8.00 \\
\midrule\multirow{8}{*}{S6} 
& AHL & 48.31 $\pm$ 30.98 & 0.49 $\pm$ 0.23 & 0.56 $\pm$ 0.20 & 0.63 $\pm$ 0.21 & 0.68 $\pm$ 0.18 & 0.38 $\pm$ 0.50 & 0.76 $\pm$ 0.19 & 1.32 $\pm$ 2.46 \\
& AHR & 42.83 $\pm$ 19.44 & 0.62 $\pm$ 0.19 & 0.66 $\pm$ 0.17 & 0.70 $\pm$ 0.20 & 0.74 $\pm$ 0.17 & 0.53 $\pm$ 0.47 & 0.82 $\pm$ 0.17 & 0.76 $\pm$ 1.25 \\
& AL & 48.63 $\pm$ 21.44 & 0.52 $\pm$ 0.22 & 0.57 $\pm$ 0.20 & 0.65 $\pm$ 0.21 & 0.69 $\pm$ 0.19 & 0.44 $\pm$ 0.48 & 0.77 $\pm$ 0.19 & 1.51 $\pm$ 2.70 \\
& AR & 28.86 $\pm$ 10.62 & 0.76 $\pm$ 0.14 & 0.79 $\pm$ 0.12 & 0.79 $\pm$ 0.17 & 0.82 $\pm$ 0.15 & 0.71 $\pm$ 0.38 & 0.89 $\pm$ 0.14 & 0.55 $\pm$ 1.03 \\
& ECL & 56.35 $\pm$ 21.61 & 0.52 $\pm$ 0.21 & 0.57 $\pm$ 0.19 & 0.66 $\pm$ 0.21 & 0.71 $\pm$ 0.18 & 0.43 $\pm$ 0.48 & 0.77 $\pm$ 0.18 & 1.55 $\pm$ 2.61 \\
& ECR & 35.72 $\pm$ 11.22 & 0.69 $\pm$ 0.16 & 0.72 $\pm$ 0.15 & 0.74 $\pm$ 0.18 & 0.77 $\pm$ 0.16 & 0.67 $\pm$ 0.39 & 0.87 $\pm$ 0.14 & 0.76 $\pm$ 1.18 \\
& PHL & 29.04 $\pm$ 16.25 & 0.44 $\pm$ 0.30 & 0.55 $\pm$ 0.24 & 0.61 $\pm$ 0.22 & 0.67 $\pm$ 0.19 & 0.34 $\pm$ 0.51 & 0.75 $\pm$ 0.19 & 1.10 $\pm$ 1.86 \\
& PHR & 38.91 $\pm$ 15.87 & 0.64 $\pm$ 0.18 & 0.68 $\pm$ 0.16 & 0.73 $\pm$ 0.19 & 0.76 $\pm$ 0.16 & 0.53 $\pm$ 0.45 & 0.81 $\pm$ 0.18 & 0.75 $\pm$ 1.43 \\
\midrule\multirow{8}{*}{S9} 
& AHL & 30.48 $\pm$ 10.93 & 0.66 $\pm$ 0.19 & 0.71 $\pm$ 0.16 & 0.80 $\pm$ 0.18 & 0.81 $\pm$ 0.16 & 0.39 $\pm$ 0.51 & 0.74 $\pm$ 0.20 & 1.22 $\pm$ 2.68 \\
& AHR & 42.39 $\pm$ 26.01 & 0.37 $\pm$ 0.29 & 0.47 $\pm$ 0.26 & 0.69 $\pm$ 0.22 & 0.71 $\pm$ 0.20 & 0.24 $\pm$ 0.52 & 0.69 $\pm$ 0.21 & 7.02 $\pm$ 15.96 \\
& AL & 37.40 $\pm$ 13.91 & 0.58 $\pm$ 0.21 & 0.63 $\pm$ 0.19 & 0.76 $\pm$ 0.19 & 0.78 $\pm$ 0.17 & 0.35 $\pm$ 0.51 & 0.72 $\pm$ 0.21 & 1.58 $\pm$ 3.21 \\
& AR & 29.71 $\pm$ 22.06 & 0.34 $\pm$ 0.35 & 0.45 $\pm$ 0.31 & 0.65 $\pm$ 0.24 & 0.69 $\pm$ 0.21 & 0.30 $\pm$ 0.54 & 0.70 $\pm$ 0.23 & 4.51 $\pm$ 8.39 \\
& ECL & 40.97 $\pm$ 20.06 & 0.53 $\pm$ 0.28 & 0.58 $\pm$ 0.26 & 0.73 $\pm$ 0.21 & 0.76 $\pm$ 0.19 & 0.38 $\pm$ 0.52 & 0.74 $\pm$ 0.21 & 1.71 $\pm$ 3.39 \\
& ECR & 43.92 $\pm$ 20.59 & 0.41 $\pm$ 0.27 & 0.48 $\pm$ 0.23 & 0.67 $\pm$ 0.23 & 0.70 $\pm$ 0.20 & 0.28 $\pm$ 0.52 & 0.70 $\pm$ 0.21 & 3.86 $\pm$ 8.00 \\
& PHL & 26.67 $\pm$ 13.45 & 0.72 $\pm$ 0.19 & 0.76 $\pm$ 0.17 & 0.83 $\pm$ 0.17 & 0.84 $\pm$ 0.15 & 0.47 $\pm$ 0.49 & 0.77 $\pm$ 0.20 & 1.07 $\pm$ 1.94 \\
& PHR & 37.42 $\pm$ 16.85 & 0.49 $\pm$ 0.27 & 0.56 $\pm$ 0.24 & 0.75 $\pm$ 0.20 & 0.77 $\pm$ 0.19 & 0.38 $\pm$ 0.50 & 0.74 $\pm$ 0.20 & 5.57 $\pm$ 11.69 \\
\bottomrule
\end{tabular}
}
\\[1ex]
\footnotesize{
    $^1$ The results of each region are averaged across all electrodes within that region.\\
    $^2$ Abbreviations for MTL Regions: AHL/AHR - Anterior Hippocampus (Left/Right), AL/AR - Amygdala (Left/Right), ECL/ECR - Entorhinal Cortex (Left/Right), PHL/PHR - Parahippocampal Gyrus (Left/Right), LR - Lateral Rhinal Area.
}
\end{table*}

The performance of NeuroFlowNet is evaluated based on its ability to generate iEEG signals from EEG signals in terms of temporal waveform, power spectral density (PSD) waveform, alpha PSD waveform and functional connection mode. The results reveal a clear pattern of region- and subject-specific variability, reflecting both the strengths and current limitations of the model.

\subsection{Temporal Waveform Comparison}

\begin{figure*}
    \centering
    \includegraphics[width=\textwidth]{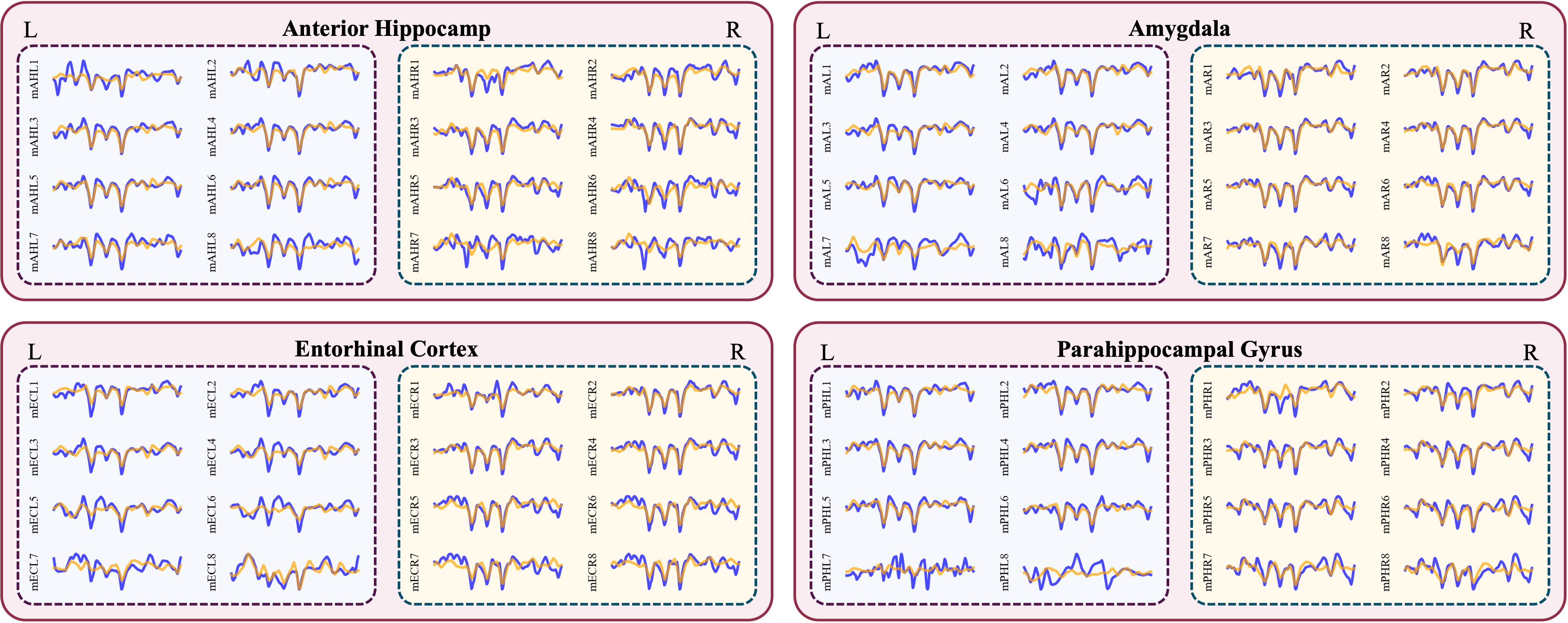}
    \caption{Exemplar generated iEEG signals (orange) overlaid with ground truth iEEG signals (blue) for various MTL subregions, specifically showing a 200ms segment from Subject S6. The figure illustrates the performance of NeuroFlowNet in generating iEEG signals from EEG data across different anatomical targets, including anterior hippocampus (AHL/AHR), amygdala (AL/AR), entorhinal cortex (ECL/ECR), and parahippocampal gyrus (PHL/PHR).}
    \label{fig:temporal_waveform_comparison}
\end{figure*}

A qualitative assessment of the model's performance is presented in Figure~\ref{fig:temporal_waveform_comparison}, which displays exemplar generated iEEG signals (orange) overlaid with the corresponding ground truth iEEG signals (blue) for a 200ms segment from Subject S6. The figure provides a visual comparison across various MTL subregions, including the anterior hippocampus (AHL/AHR), amygdala (AL/AR), entorhinal cortex (ECL/ECR), and parahippocampal gyrus (PHL/PHR). As showed in Figure~\ref{fig:temporal_waveform_comparison}, the generated waveforms demonstrate a high degree of fidelity to the ground truth signals, successfully capturing the principal morphological characteristics and temporal dynamics. The model effectively replicates both the slower, underlying oscillations and the relatively faster fluctuations that remain observable under the 200~Hz bandwidth. The alignment is particularly strong in regions such as the amygdala and the parahippocampal gyrus, where the generated signals closely trace the contours of the ground truth waveforms in terms of phase and relative amplitude.

In more structurally complex or deeper regions, such as the anterior hippocampus and entorhinal cortex, the model still captures the general trend of the signal but occasionally shows minor discrepancies in peak amplitude or precise temporal alignment. For instance, some of the fast transients in the ground truth signal are slightly smoothed or have a marginally lower amplitude in the generated counterpart. Nevertheless, the overall correspondence remains robust, underscoring NeuroFlowNet's capability to learn the complex conditional mapping from non-invasive EEG to invasive iEEG signals across multiple anatomical targets. This visual evidence corroborates that the model generates temporally coherent and physiologically plausible iEEG waveforms.

To quantify the temporal waveform performance of the model, Tables~\ref{tab:performance_results_trial} and~\ref{tab:performance_results_trial_session} report the mean absolute error (MAE), Pearson correlation coefficient (Corr), and cosine similarity across multiple MTL subregions for subjects S1, S6, and S9 under the cross-trial and cross-session settings, respectively. Averaged across all listed regions, the cross-trial setting yields an MAE of 39.93~$\mu$V, a correlation of 0.50, and a cosine similarity of 0.58, while the cross-session setting yields 47.22~$\mu$V, 0.54, and 0.61. Therefore, moving from unseen trials to unseen sessions does not cause a systematic collapse of waveform similarity; instead, the temporal metrics remain in the same range, indicating that NeuroFlowNet captures subject-specific EEG-to-iEEG mappings that are not restricted to a single recording session.

At the regional level, the same broad performance pattern is preserved in both settings. More stable reconstructions are typically observed in regions such as S6-AR and S9-PHL in the cross-trial setting and S6-AR, S6-ECR, S1-LR, and S1-PHL in the cross-session setting, all of which show temporal correlations above 0.69 and, in some cases, above 0.80. In contrast, deeper or more difficult targets, such as S1-AHL and especially S1-PHR, remain challenging across both settings. This consistency suggests that the dominant source of variability is region- and subject-dependent observability rather than a simple overfitting to particular trials.

Notably, MAE is more sensitive to session-level amplitude shifts than correlation-based metrics. For example, in subject S1, the average MAE increases from 47.43~$\mu$V in the cross-trial setting to 70.21~$\mu$V in the cross-session setting, whereas the average temporal correlation simultaneously increases from 0.42 to 0.51. This divergence indicates that session-wise changes in signal scale or baseline can affect amplitude fidelity even when waveform morphology remains aligned. Taken together, the temporal results support a robustness claim that is appropriately limited: NeuroFlowNet generalizes to unseen sessions within a subject, but amplitude calibration is still more fragile than waveform-shape recovery.

\subsection{Power Spectral Density Comparison}
\begin{figure*}
    \centering
    \includegraphics[width=\textwidth]{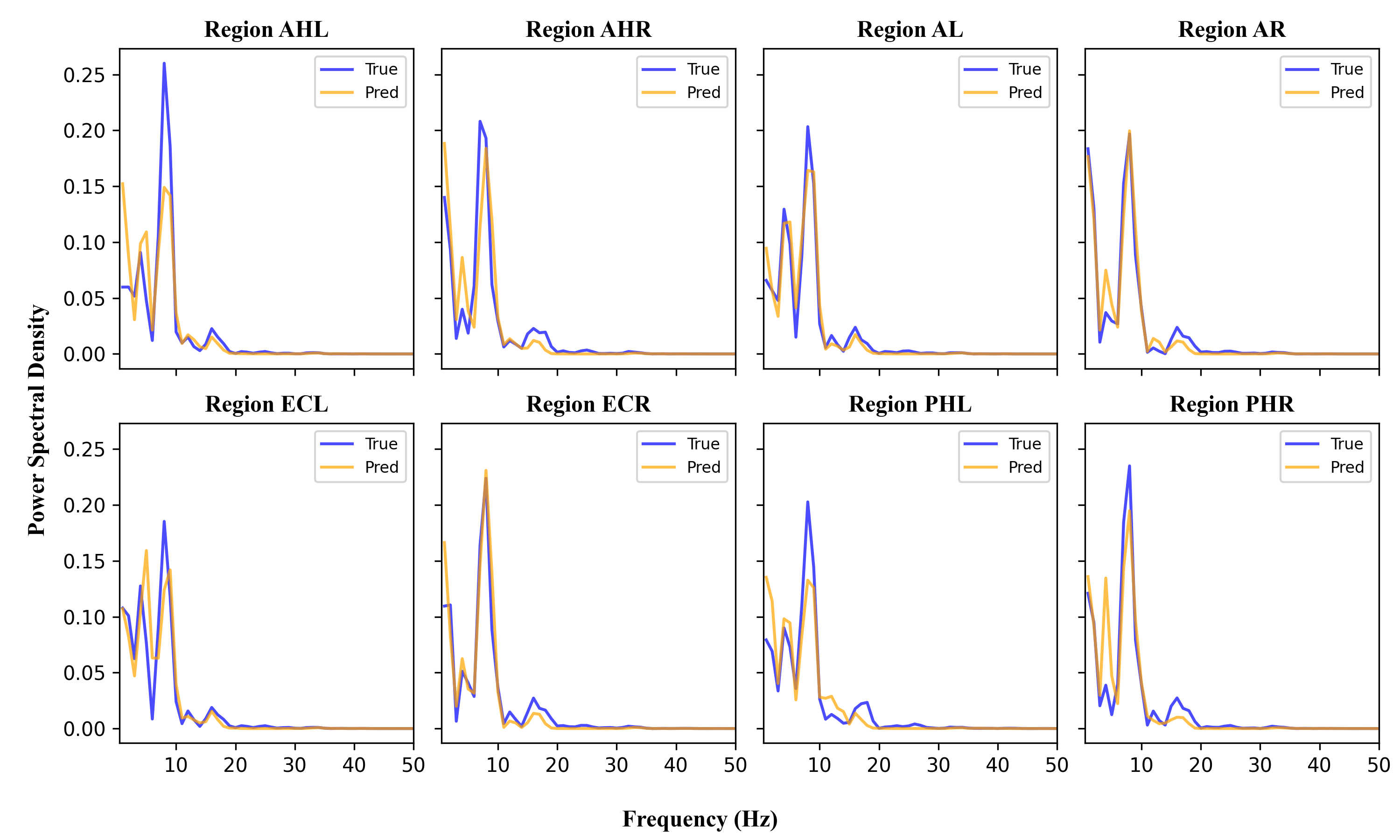}
    \caption{Comparison of power spectral density (PSD) between ground truth iEEG signals (blue) and generated iEEG signals (orange) for a 200 ms segment from Subject S6. The figure shows the average PSD across all electrodes within each MTL subregion, including anterior hippocampus (AHL/AHR), amygdala (AL/AR), entorhinal cortex (ECL/ECR), and parahippocampal gyrus (PHL/PHR).}
\label{fig:psd_comparison}
\end{figure*}

To evaluate the frequency reproduction capability within the preserved bandwidth of NeuroFlowNet, Figure~\ref{fig:psd_comparison} shows a comparison of the PSD waveforms of the generated signals (orange) and the actual signals (blue) for eight MTL subregions of subject S6. The PSD waveform for each region is the average of all electrodes in that region. The analysis reveals that NeuroFlowNet successfully captures the salient spectral characteristics of the native iEEG signals. Across all depicted regions, the generated PSDs closely mirror the ground truth within the preserved bandwidth, particularly in replicating the prominent power peak in the alpha band (8-13 Hz) and a secondary peak in the theta band (4-8 Hz)~\cite{herweg2020theta,jensen2010shaping}. These frequency bands are physiologically critical in the MTL; theta oscillations are integral to memory encoding and retrieval, while alpha rhythms are associated with functional inhibition and attentional modulation. The model's accurate generation of these peaks indicates that it learns to infer the state of these underlying cognitive processes from the conditioning EEG data. Furthermore, the model correctly reproduces the aperiodic, 1/f-like decay in power at higher frequencies, a hallmark of background neural activity~\cite{he2014scale,kramer20231}. This demonstrates a robust generalization beyond just the dominant oscillatory components.

\begin{figure*}
    \centering
    \includegraphics[width=\textwidth]{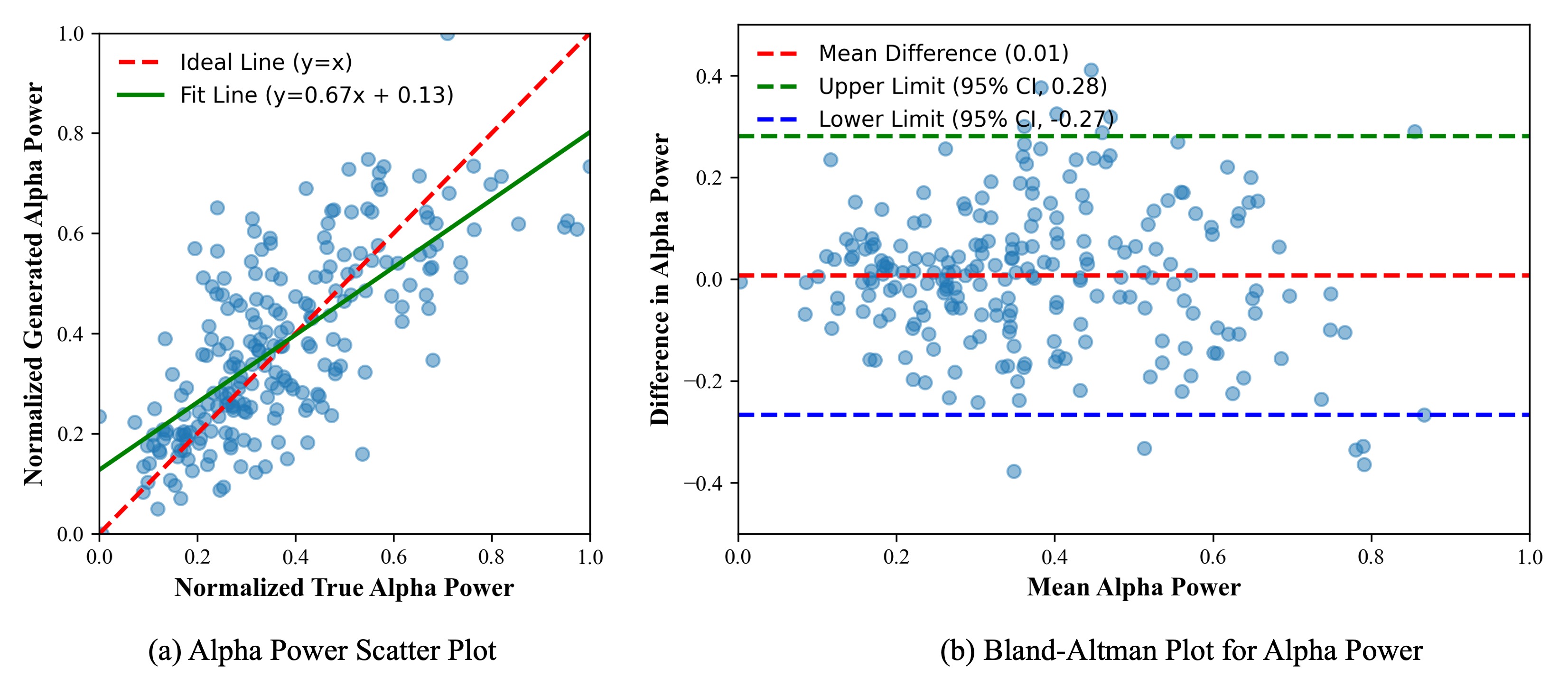}
    \caption{(a) Scatter plot of normalized true alpha power versus normalized generated alpha power. The red dashed line represents the ideal line (y = x), while the green solid line indicates the linear fit (y = 0.67x + 0.13). (b) Bland-Altman plot showing the difference in alpha power against the mean alpha power. The red dashed line indicates the mean difference (0.01), and the green and blue dashed lines represent the upper (95\% CI, 0.28) and lower (95\% CI, -0.27) limits of agreement, respectively.}
    \label{fig:alpha_psd_comparison}
\end{figure*}

To further scrutinize the model's performance on the functionally significant alpha band, Figure~\ref{fig:alpha_psd_comparison} provides a detailed analysis of generated versus true alpha power. The scatter plot (Figure~\ref{fig:alpha_psd_comparison}a) shows a strong positive correlation, with the linear fit line (y=0.67x+0.13) closely following the ideal line (y=x). This indicates that the model reliably captures the relative changes in alpha power. The slope of less than one suggests a slight regression to the mean, where the model may temper the most extreme high or low power values, a common characteristic when inferring high-resolution signals from lower signal-to-noise data. The Bland-Altman plot (Figure~\ref{fig:alpha_psd_comparison}b) reinforces this finding by showing that the mean difference between true and generated alpha power is exceptionally low (0.01), indicating no systematic bias in the model's predictions. The data points are evenly scattered within the 95\% limits of agreement, confirming that while individual predictions have some variance, the model's estimations are, on average, highly accurate. Collectively, the spectral analysis demonstrates that NeuroFlowNet generates iEEG signals that are not only temporally plausible but also spectrally realistic, preserving the key neurophysiological frequency fingerprints essential for downstream analysis.

To quantify the frequency-domain reproduction capability of the model, Tables~\ref{tab:performance_results_trial} and~\ref{tab:performance_results_trial_session} summarize the 0.5-50~Hz PSD waveform metrics (Corr and cosine similarity) and the 8-13~Hz $\alpha$ PSD waveform metrics (Corr, cosine similarity, and mean absolute percentage error (MAPE)) for multiple MTL subregions of subjects S1, S6, and S9. Across all listed regions, the average PSD correlation/cosine pair changes only slightly from 0.698/0.733 in the cross-trial setting to 0.725/0.756 in the cross-session setting. Likewise, the average alpha-band cosine similarity remains nearly unchanged (0.744 vs. 0.751), and the average alpha power MAPE stays low in both settings (2.43\% vs. 2.78\%). These results indicate that the low-/mid-frequency spectral fingerprints learned by NeuroFlowNet are stable not only across unseen trials but also across unseen sessions from the same subject.

At the region level, the spectral performance is again heterogeneous but consistent with the temporal observations. In the broad frequency range (0.5-50~Hz), PSD correlations remain high in many regions, from 0.53 $\pm$ 0.25 (PHR, S1) to 0.84 $\pm$ 0.16 (PHL, S9) in the cross-trial setting and up to 0.93 $\pm$ 0.12 in S1-LR and S1-PHL in the cross-session setting. Most PSD cosine similarities exceed 0.70, suggesting that the overall spectral profile is preserved even when the exact waveform amplitude is imperfect. In the alpha band, cosine similarities remain stable, whereas correlations fluctuate more strongly, indicating that the model is better at preserving relative alpha-band shape and power than exact segment-wise alpha fluctuations.

The main degradation under the session-level split is concentrated in a small number of difficult regions, especially S1-PHR and S1-ECL, where both temporal and alpha-band errors are larger. Even so, the overall session-level spectral metrics remain comparable to the trial-level metrics. This is important for the robustness discussion: the model is not merely memorizing trial-specific spectral statistics from sessions seen during training, but instead preserves meaningful band-limited frequency structure when transferred to held-out sessions. Such behavior is particularly relevant for downstream neurophysiological analyses that rely more on stable band-power organization than on exact pointwise waveform matching~\cite{assenza2017oscillatory,uhlhaas2006neural,zhou2021relative}.

\subsection{Inter-Channel Correlation Comparison}
\label{iccc}
\begin{figure*}
    \centering
    \includegraphics[width=\textwidth]{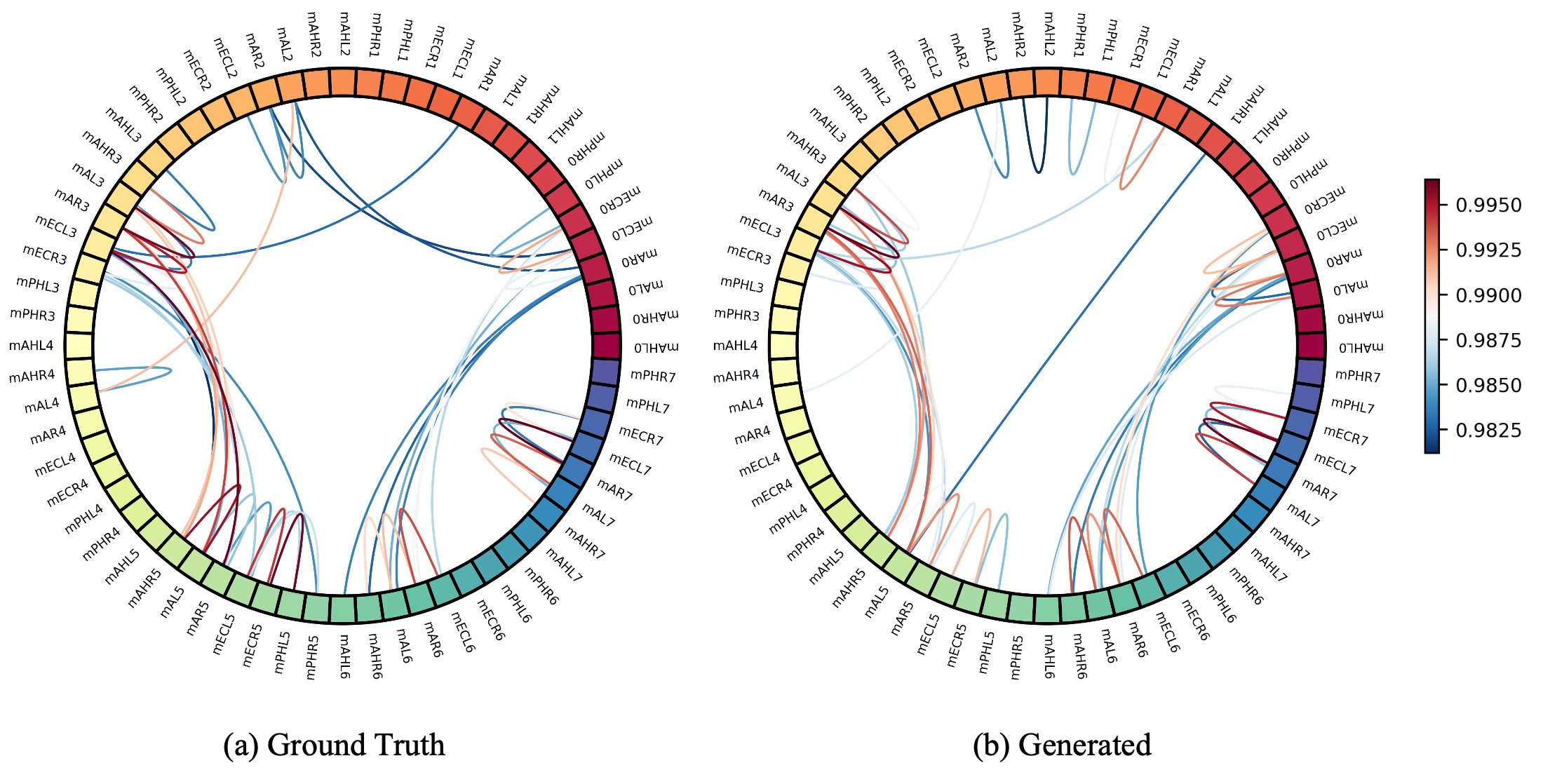}
    \caption{Comparison of inter-channel correlation. Circular graphs show the connectivity patterns of (a) the original iEEG signals and (b) the generated iEEG signals. Both plots illustrate a 200 ms segment from Subject S6. Pairwise Pearson correlation coefficients are computed between all channels, and only the top 50 strongest absolute correlations are visualized for clarity.}
    \label{fig:conn_comparison}
\end{figure*}

Beyond generating realistic single-channel waveforms, a crucial test for the model is its ability to capture the complex functional relationships between different neural populations. These relationships are reflected in the inter-channel correlation structure of the iEEG signals. Figure~\ref{fig:conn_comparison} presents a qualitative comparison of these correlation patterns, displaying the 50 strongest pairwise Pearson correlations for both the ground truth (a) and generated (b) iEEG signals from a 200ms segment of Subject S6.

The circular plots reveal a close correspondence between the ground truth and the generated data. The model reproduces salient features of the correlation topology visible in this band-limited representation. Key patterns, such as strong correlations between homologous regions in the left and right hemispheres (e.g., between channels in AHL and AHR), are clearly preserved.

Furthermore, the model captures strong correlations between anatomically and functionally related regions within the same hemisphere, such as the entorhinal cortex and hippocampus. The visual similarity of the two plots—from the dense connections within the parahippocampal gyrus to the sparser but strong links spanning distant regions—suggests that NeuroFlowNet is not merely generating a collection of independent, realistic time series, but is also retaining cross-channel covariance patterns observable in the reconstructed signals. Since Pearson correlation is primarily sensitive to shared low-/mid-frequency fluctuations, this comparison is most appropriately interpreted as evidence that the model preserves one important aspect of MTL dependency structure within the retained bandwidth, rather than fully recovering the underlying neural or network dynamics.

\subsection{Conditional Sampling and Uncertainty Analysis}
\label{sec:conditional_sampling_uncertainty}
To directly examine whether NeuroFlowNet behaves as a stochastic conditional generator rather than a deterministic reconstructor, we performed a conditional sampling and uncertainty analysis on the validation set. For each validation EEG segment, we repeatedly drew $K=20$ independent latent samples from the base Gaussian distribution and generated 20 corresponding iEEG reconstructions using the same trained subject-specific model and the same preprocessing/postprocessing pipeline. This protocol keeps the conditioning EEG fixed and changes only the latent variable, so any variation among the generated outputs reflects the learned conditional distribution $p(X_{iEEG}\mid X_{EEG})$.

For each segment, we computed the sample mean, sample variance, and empirical 5th--95th percentile interval across the 20 conditional samples. Conditional sample diversity was quantified in two complementary ways: (i) the average pairwise RMSE between generated waveforms in the time domain, and (ii) the average pairwise Frobenius distance between inter-channel correlation matrices computed from different samples. We further evaluated the 90\% predictive interval coverage probability (PICP), mean prediction interval width (MPIW), MAE and temporal correlation of the sample mean, and uncertainty-error consistency. The latter was measured as the Spearman correlation between segment-level predictive uncertainty (mean sample variance across channels and time) and segment-level reconstruction error (MAE of the sample mean).

\begin{figure*}[t]
    \centering
    \includegraphics[width=\textwidth]{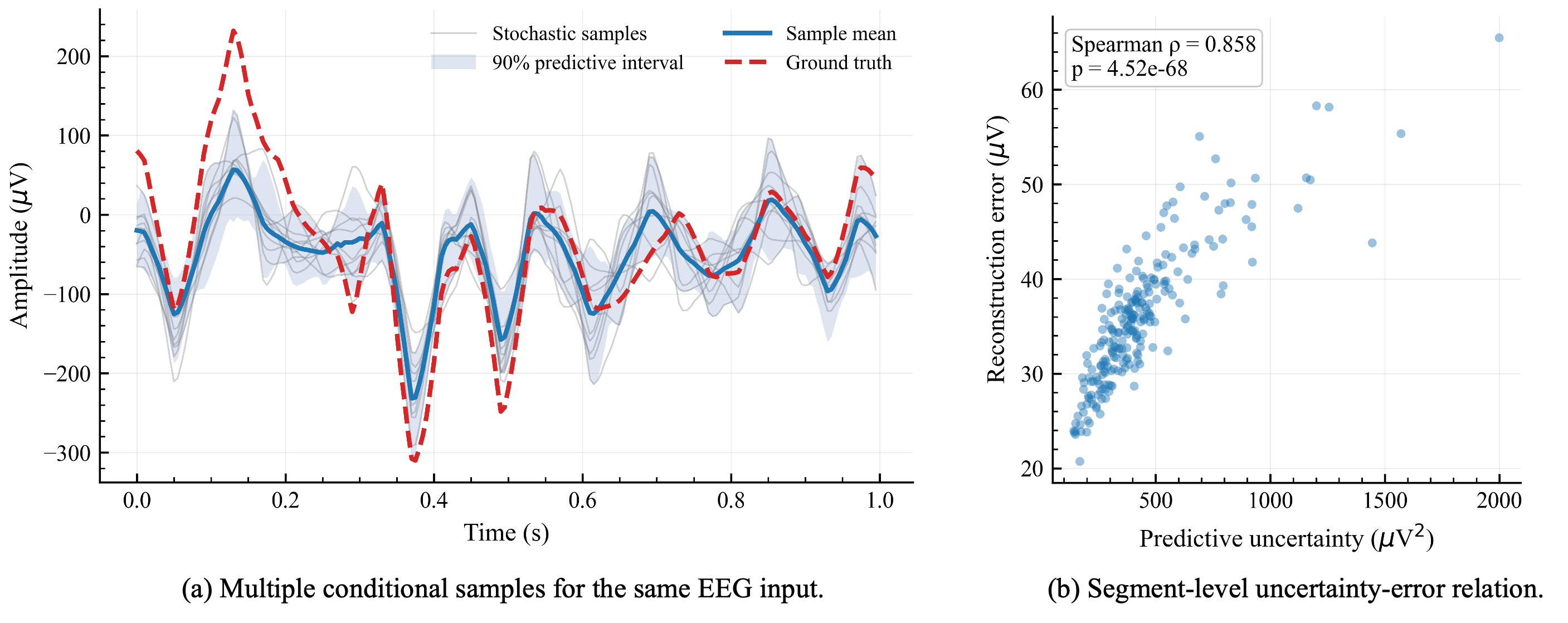}
    \caption{
    Representative conditional sampling and uncertainty analysis for Subject S6.
    (a) For a fixed scalp EEG segment, $K=20$ independent latent samples generate multiple iEEG trajectories for channel mAHL1. The solid line denotes the sample mean, the shaded area denotes the empirical 90\% predictive interval, and the dashed line denotes the ground-truth iEEG.
    (b) Across all validation segments of S6, each point represents one segment. Predictive uncertainty is defined as the mean sample variance across channels and time, and reconstruction error is defined as the MAE between the sample mean and the ground truth. Higher predictive uncertainty is strongly associated with larger reconstruction error ($\rho = 0.858$, $p = 4.52 \times 10^{-68}$).
    }
    \label{fig:conditional_sampling_uncertainty}
\end{figure*}

\begin{table*}[t]
\centering
\caption{Conditional sampling and uncertainty metrics on the validation set.}
\label{tab:conditional_sampling_uncertainty}
\resizebox{\textwidth}{!}{
\begin{tabular}{lcccccccccc}
\toprule
\textbf{Subject} & \textbf{Segments} & \textbf{Channels} & \textbf{Diversity RMSE} & \textbf{Corr. diversity} & \textbf{PICP$_{90}$} & \textbf{MPIW} & \textbf{Mean variance} & \textbf{Mean MAE} & \textbf{Mean Corr.} & \textbf{Unc.-error $\rho$} \\
 & & & \textbf{($\mu$V)} & \textbf{(Frobenius)} & & \textbf{($\mu$V)} & \textbf{($\mu$V$^2$)} & \textbf{($\mu$V)} & & \\
\midrule
S1 & 112 & 48 & 35.59 & 7.55 & 0.381 & 62.00 & 841.51 & 49.37 & 0.490 & 0.958 ($p=2.03\times10^{-61}$) \\
S6 & 231 & 64 & 28.56 & 8.64 & 0.432 & 51.57 & 434.66 & 35.78 & 0.599 & 0.858 ($p=4.52\times10^{-68}$) \\
S9 & 70 & 64 & 23.48 & 7.99 & 0.382 & 42.94 & 286.27 & 34.01 & 0.572 & 0.878 ($p=1.92\times10^{-23}$) \\
\bottomrule
\end{tabular}
}
\\[1ex]
\footnotesize{
    $K=20$ conditional samples were generated for each validation EEG segment. Diversity RMSE is the average pairwise RMSE among conditional samples. Correlation diversity is the average pairwise Frobenius distance between inter-channel correlation matrices. PICP$_{90}$ and MPIW are computed from the empirical 5th--95th percentile interval. Mean Corr. denotes the temporal correlation between the sample mean reconstruction and ground-truth iEEG.
}
\end{table*}

The analysis provides direct empirical evidence that NeuroFlowNet generates non-degenerate conditional samples. Across subjects, the average pairwise time-domain diversity ranged from 23.48 to 35.59~$\mu$V, and the pairwise diversity of inter-channel correlation matrices ranged from 7.55 to 8.64 in Frobenius norm (Table~\ref{tab:conditional_sampling_uncertainty}). Thus, repeated sampling under an identical EEG condition does not collapse to a single deterministic trace; instead, the model produces a family of EEG-consistent iEEG trajectories with measurable waveform-level and connectivity-level variation. The representative example in Figure~\ref{fig:conditional_sampling_uncertainty}(a) illustrates this behavior visually: individual stochastic reconstructions fluctuate around the sample mean while retaining the main morphology of the ground-truth signal.

The uncertainty estimates also showed a consistent relationship with reconstruction difficulty. Segment-level predictive variance was strongly and positively correlated with segment-level MAE for all three subjects (S1: $\rho=0.958$, $p=2.03\times10^{-61}$; S6: $\rho=0.858$, $p=4.52\times10^{-68}$; S9: $\rho=0.878$, $p=1.92\times10^{-23}$). This indicates that the stochastic spread of conditional samples is not arbitrary noise; segments for which the model is more uncertain also tend to be segments with larger reconstruction error. Region-level results showed the same pattern. For example, the S1 PHR region exhibited both the largest uncertainty (mean variance: 2580.25~$\mu$V$^2$) and the largest sample-mean error (84.46~$\mu$V), whereas lower-error regions such as S1 PHL and S6/S9 PHL showed substantially smaller uncertainty.

At the same time, the empirical predictive intervals were not fully calibrated: PICP$_{90}$ ranged from 0.381 to 0.432, well below the nominal 0.90 level. Therefore, these results should not be interpreted as evidence of a calibrated Bayesian posterior interval. Rather, they support the more specific conclusion that NeuroFlowNet exhibits genuine conditional sample diversity and that its sample variance provides an informative, although under-dispersed, uncertainty signal. This directly strengthens the generative interpretation of the model while keeping the scope of the claim limited to stochastic behavior and uncertainty-error consistency under the present subject-specific reconstruction protocol.

\subsection{Functional Connectivity Fidelity Comparison with Regression Baselines}

\begin{table*}[h]
\centering
\caption{Functional connectivity fidelity (inter-channel correlation structure) comparison between NeuroFlowNet and regression baselines.}
\label{tab:connectivity_baselines_horizontal}
\begin{tabular}{llllll}
\toprule
\textbf{Matrix} & \textbf{Linear (1$\times$1)} & \textbf{Shallow CNN} & \textbf{1D U-Net} & \textbf{Tiny Transformer} & \textbf{NeuroFlowNet (Ours)} \\
\midrule
$\mathcal{E}_{\mathrm{MAE}}(\mathbf{R})$ & 0.426 $\pm$ 0.064 & 0.318 $\pm$ 0.032 & 0.302 $\pm$ 0.013 & 0.360 $\pm$ 0.080 & \textbf{0.288 $\pm$ 0.042}\\
$\mathcal{E}_{F}(\mathbf{R})$            & 30.45 $\pm$ 6.45 & 23.37 $\pm$ 3.99 & 22.08 $\pm$ 2.57 & 26.04 $\pm$ 6.86 & \textbf{21.61 $\pm$ 4.69}\\
\bottomrule
\end{tabular}
\footnotesize
Note: $\mathbf{R}$ denotes the inter-channel Pearson correlation matrix computed over time for each 1-second segment. $\mathcal{E}_{\mathrm{MAE}}(\mathbf{R}) = \frac{1}{C^2}\lVert \mathbf{R}^{\text{pred}}-\mathbf{R}^{\text{true}}\rVert_{1}$ and $\mathcal{E}_{F}(\mathbf{R})=\lVert \mathbf{R}^{\text{pred}}-\mathbf{R}^{\text{true}}\rVert_{F}$, averaged within each subject and then across subjects, reported as mean $\pm$ std.
\end{table*}

As emphasized above, an important requirement for non-invasive iEEG reconstruction is not only waveform plausibility at each channel, but also the preservation of inter-channel correlation structure that reflects MTL functional connectivity. In practice, many downstream analyses in cognitive and clinical neuroscience rely on network-level signatures (e.g., co-fluctuations across contacts/regions), and a model that produces visually plausible traces yet distorts cross-channel dependencies may be of limited utility for connectivity-driven interpretation. Therefore, we complement our within-model connectivity demonstration with a controlled comparison to representative deterministic regression baselines under the protocol defined in Section~\ref{sec:connectivity_baseline_setup}.

Table~\ref{tab:connectivity_baselines_horizontal} summarizes the correlation-structure fidelity across models. NeuroFlowNet achieves the lowest errors among the compared methods, with $\mathcal{E}_{\mathrm{MAE}}(\mathbf{R})=0.288\pm0.042$ and $\mathcal{E}_{F}(\mathbf{R})=21.61\pm4.69$. Compared with the pointwise linear baseline, NeuroFlowNet reduces correlation-matrix MAE from $0.426\pm0.064$ to $0.288\pm0.042$ (a relative reduction of $\sim$32\%), and reduces Frobenius error from $30.45\pm6.45$ to $21.61\pm4.69$ (a relative reduction of $\sim$29\%). This indicates that a simple per-time linear mapping is insufficient to recover the cross-channel dependency patterns expressed in these band-limited iEEG signals, whereas NeuroFlowNet better preserves the corresponding correlation structure under the same training data regime.

Beyond the linear baseline, NeuroFlowNet also consistently outperforms the stronger nonlinear regressors included in this comparison (Shallow CNN, 1D U-Net, and Tiny Transformer). While these deterministic models can learn nonlinear mappings and often generate temporally smooth outputs, they still tend to emphasize pointwise signal reconstruction objectives, which can inadvertently \emph{attenuate} or \emph{homogenize} cross-channel relationships—especially when the inverse mapping from scalp EEG to deep iEEG is underconstrained. In such ill-posed settings, a deterministic regressor is incentivized to output a ``central'' solution that reduces average prediction error across trials, which may appear reasonable at the level of individual traces but can shrink trial-to-trial covariance and distort the correlation matrix. The observed gap in $\mathcal{E}_{\mathrm{MAE}}(\mathbf{R})$ and $\mathcal{E}_{F}(\mathbf{R})$ therefore suggests that NeuroFlowNet better preserves the correlation structure most visible to the present Pearson-correlation analysis, aligning with the qualitative similarity of connectivity graphs shown in Figure~\ref{fig:conn_comparison}. Because this metric is dominated by shared low-/mid-frequency fluctuations and by the bandwidth retained after preprocessing, the result is best understood as improved preservation of correlation structure within that reconstructed frequency range.

\subsection{Architectural Interpretation and Scope of Connectivity Preservation}

One conceptual explanation for this advantage lies in the inductive biases of conditional normalizing flows and the specific architectural design of NeuroFlowNet. First, CNF optimizes exact conditional likelihood, enabling the model to represent the \emph{conditional distribution} of iEEG given EEG rather than collapsing all variability into a single deterministic output. Conceptually, this is consistent with the non-uniqueness of EEG inverse problems: multiple intracranial configurations can give rise to similar scalp observations due to volume conduction and source mixing. By introducing a latent variable and learning an invertible mapping, NeuroFlowNet can in principle retain residual degrees of freedom that are not fully determined by EEG, instead of forcing them into a mean prediction. This provides a plausible account of why the model may better preserve trial-level co-variations across channels, which then manifest as more faithful correlation matrices, although the specific physiological meaning of the latent representation is not established by the present experiments.

The invertible $1\times1$ convolution in each flow step provides an explicit channel-mixing mechanism. Unlike standard convolutions that may learn channel interactions implicitly (and potentially inconsistently across layers), the invertible mixing operation encourages structured information exchange across channels at every temporal location while maintaining reversibility. This is particularly relevant in multi-channel iEEG, where functional connectivity is encoded in coordinated activity patterns across contacts/regions. By repeatedly mixing channels through invertible transforms, the model can more effectively capture and propagate global cross-channel dependencies throughout the flow, which can improve the fidelity of reconstructed correlation structure.

In addition, the multi-scale architecture further supports structure preservation by decomposing the learning problem across resolutions. Fine scales focus on local temporal detail, whereas coarser scales capture more global, slowly varying patterns that often drive connectivity estimates. Since correlation matrices integrate information over time, errors in low-frequency or large-scale components can disproportionately affect connectivity fidelity. The hierarchical split-and-propagate design allows NeuroFlowNet to allocate capacity to both local waveform characteristics and global coordination patterns, which may explain its stronger performance in $\mathcal{E}_{\mathrm{MAE}}(\mathbf{R})$ and $\mathcal{E}_{F}(\mathbf{R})$.

The role of the multi-head self-attention (MHSA) module should also be interpreted in this context. MHSA is not introduced primarily as a smoothing operator; rather, its main function is to model long-range temporal dependencies that are difficult to capture with local convolutions alone. In EEG-to-iEEG reconstruction, such dependencies can link temporally separated events, phase relationships, and slowly evolving contextual states that shape the instantaneous waveform. Any apparent smoothing in the generated traces is therefore better understood as a by-product of learning temporally coherent conditional structure under a band-limited setting, not as the principal purpose of attention itself.

Relatedly, the different architectural components likely contribute unevenly to waveform reconstruction and connectivity preservation. The finer scales and local convolutional operations are expected to contribute more directly to pointwise waveform morphology, including peak shape, local oscillatory detail, and short-lag temporal alignment. By contrast, the coarser scales, MHSA, and repeated channel-mixing operations are more naturally aligned with preserving slowly varying coordination patterns across time and channels, which are the main drivers of the inter-channel correlation analysis used here. Because the present experiments were not designed as a strict component-wise ablation, this interpretation should be viewed as a mechanistic explanation of the observed performance profile rather than a definitive quantitative attribution. Nevertheless, it provides a coherent account of why NeuroFlowNet can simultaneously improve single-channel realism and better retain band-limited connectivity structure.

These findings should be interpreted within the subject-specific scope of this dataset. The present comparison demonstrates that, under identical within-subject training conditions, NeuroFlowNet yields improved fidelity of band-limited inter-channel correlation structure relative to representative deterministic regressors. Together with the conditional sampling results in Section~\ref{sec:conditional_sampling_uncertainty}, this supports a more specific and empirically grounded claim: within the current protocol, the CNF formulation improves structure-preserving reconstruction over deterministic regressors while also producing observable stochastic conditional samples with informative uncertainty behavior. A protocol-matched comparison with other generative families, such as diffusion models or GANs, remains an important direction for future work. Future studies will also extend these connectivity-driven evaluations to broader cohorts and more diverse brain states, and will assess whether this correlation-structure advantage translates into improved performance on downstream tasks that explicitly rely on network-level features (e.g., connectivity-based decoding or seizure network characterization).

\subsection{Computational latency and real-time feasibility}
Since NeuroFlowNet relies on reversible transformations, we further evaluated the inference latency to assess its feasibility for real-time monitoring. The latency was measured using the final model configuration selected in this study, i.e., \(N_S = 2\), \(N_{\mathrm{steps}} = 6\), and \(C_h = 128\). For each test, a 1-s EEG segment sampled at 200 Hz was used as the conditioning input, and the model generated the corresponding iEEG segment under batch size 1. All measurements were performed in evaluation mode with gradient computation disabled, and the first 50 runs were discarded as warm-up. On a single NVIDIA RTX 4080 GPU, the average inference latency was \(9.536 \pm 2.449~\mathrm{ms}\) per 1-s segment, with a median latency of \(8.724~\mathrm{ms}\) and a 95th-percentile latency of \(14.253~\mathrm{ms}\). The corresponding real-time factor was 0.00954, indicating that the inverse reversible transformations required less than \(1\%\) of the input window duration. Even the maximum observed latency, \(26.609~\mathrm{ms}\), remained far below the 1000-ms window length. These results suggest that NeuroFlowNet has practical potential for real-time or near-real-time monitoring.

The computational cost of the reversible transformations is mainly determined by the repeated evaluations of the conditional network \(g_{\mathrm{cond}}\) in the affine coupling layers and the inverse \(1 \times 1\) convolutions. Importantly, during inference, the model does not require iterative optimization; iEEG generation is achieved by a single inverse pass through the learned flow. Therefore, although the reversible architecture introduces additional transformation steps compared with simple deterministic regression models, its inference process remains deterministic in computational schedule and can be efficiently accelerated on modern GPUs. Future work will further optimize the model for deployment on low-power edge devices and streaming EEG systems.

\subsection{Discussion of Physiological Determinants in Regional Variability}
The observed performance heterogeneity across MTL subregions can be partially attributed to neuroanatomical and physiological factors. In terms of structural complexity, the anterior hippocampus (AHL/AHR) is characterized by intricate laminar organization and high neuronal density, resulting in complex and non-linear iEEG signal patterns~\cite{li2018optimal}. This complexity likely exceeds the current representational capacity of NeuroFlowNet's conditional normalizing flow, contributing to reduced fidelity in this region~\cite{winkler2019learning}. Besides, deeply situated structures like AHL experience greater signal attenuation in EEG recordings due to intervening tissues (cerebrospinal fluid, skull), reducing the signal-to-noise ratio of the conditioning data and complicating accurate inference of iEEG dynamics~\cite{hnazaee2020localization}. In contrast, more superficial regions such as the parahippocampal gyrus benefit from proximity to the cortical surface, aiding model performance~\cite{vlcek2020mapping}. Additionally, regions with strong cortico-MTL connectivity, such as the entorhinal cortex (ECL/ECR), may exhibit iEEG dynamics that are more inferable from scalp-recorded potentials, aligning with their relatively high temporal and spectral performance~\cite{yang2025enhanced}.

\subsection{Discussion of Bandwidth and Denoising Limitations}
\label{sec:bandwidth_limit}

A key preprocessing step in this work is downsampling iEEG from 2~kHz to 200~Hz to match scalp EEG and to enforce an aligned temporal grid for conditional modeling. While this alignment is important for stable supervised training and fair cross-modal comparison, it imposes an explicit bandwidth constraint: after downsampling, the effective Nyquist frequency is 100~Hz, and spectral content above 100~Hz cannot be represented. In addition, the required anti-aliasing low-pass filter applied before decimation/resampling can attenuate components near the cutoff, which may further reduce the sharpness of rapid fluctuations even below 100~Hz.

Another related preprocessing limitation is the use of wavelet denoising before model training. Although denoising was applied to suppress broadband noise and improve signal quality, its effect on transient waveform morphology was not directly evaluated in the present study. In particular, the denoising procedure may smooth sharp transient components or attenuate part of the faster fluctuations. Therefore, the waveform comparisons reported in this work should be interpreted as reconstruction performance relative to the denoised and band-limited target iEEG signals, rather than as evidence that all transient morphological features in the original broadband iEEG recordings were preserved.

This limitation is particularly relevant for intracranial biomarkers that are typically studied in higher frequency ranges, such as high-gamma activity and high-frequency oscillations, as well as very fast transient events. These phenomena are not the target of the present reconstruction setting and cannot be faithfully assessed under a 200~Hz sampling pipeline. Accordingly, the term ``high-fidelity'' in this study refers to fidelity within the preserved bandwidth of the preprocessing chain and relative to the denoised target signals, rather than broadband fidelity across the full spectrum available in the original 2~kHz iEEG.

Consistent with this scope, our quantitative frequency-domain evaluations are restricted to low-/mid-frequency bands, specifically PSD similarity in 0.5--50~Hz and alpha-band (8--13~Hz) characteristics (Tables~\ref{tab:performance_results_trial} and~\ref{tab:performance_results_trial_session}). Moreover, our functional-connectivity analyses rely on inter-channel correlation matrices computed from these preprocessed time series, and our conclusions are therefore confined to (i) temporal waveform plausibility at 200~Hz resolution, (ii) preservation of low-/mid-frequency spectral fingerprints, and (iii) recovery of inter-channel dependency structure within this bandwidth.

Future work will extend the framework toward higher-frequency intracranial signatures by (i) training and evaluating at higher target sampling rates, such as 500~Hz or 1~kHz, when synchronized EEG-iEEG data of sufficient quality are available, and/or (ii) adopting multi-resolution modeling strategies in which low-frequency components are reconstructed at 200~Hz for cross-modal alignment while higher-frequency components are modeled with an additional dedicated branch or a hierarchical refinement stage. Future studies should also include a systematic raw-versus-denoised comparison to quantify how preprocessing affects transient morphology before model training. These directions would enable a more comprehensive assessment of whether non-invasive EEG can support reconstruction of clinically and cognitively relevant high-frequency iEEG biomarkers beyond the present denoised and band-limited setting.

\subsection{Discussion of Physiological Interpretation of the Latent Space Representation $Z$}
To further strengthen the interpretability of NeuroFlowNet, we discuss the potential physiological meaning of the learned latent representation $Z$. At the present stage, the physiological interpretation of $Z$ is best viewed as a source of working hypotheses about the residual intracranial dynamics captured by the model, rather than as a direct identification of specific physiological processes. In conditional normalizing flows, $Z$ is obtained through an invertible transformation $Z=f(X_t,X_c)$, where $X_t$ and $X_c$ denote the target iEEG and conditioning scalp EEG, respectively. Due to the bijective nature of the flow mapping, $Z$ should not be interpreted as a single physiological ``source'' or a direct anatomical generator in the classical sense. Instead, we view $Z$ as a compact parametrization of the residual degrees of freedom in intracranial activity that are not uniquely constrained by scalp EEG, i.e., a principled representation of the conditional uncertainty in $p(X_{iEEG}\mid X_{EEG})$~\cite{papamakarios2021normalizing}.

From a neurophysiological perspective, this residual component can plausibly reflect multiple factors. First, it may capture trial-to-trial variability of local circuit dynamics within the MTL that is only weakly expressed on the scalp due to volume conduction, source mixing, and limited signal-to-noise ratio~\cite{asadzadeh2020systematic,zorzos2021advances,piastra2021comprehensive}. Second, it may encode micro-scale and transient activity (e.g., sharp fluctuations and other local population events) that substantially shapes iEEG waveforms but is attenuated or spatially smeared in scalp recordings~\cite{piastra2021comprehensive,afnan2024eeg,liu2022consensus}. Third, it may represent state-dependent stochastic modulations of ongoing rhythms, where fluctuations in cognitive state or vigilance alter intracranial oscillatory patterns without being fully predictable from scalp EEG alone~\cite{pershin2023vigilance}. Importantly, this interpretation is consistent with the well-known non-uniqueness of EEG inverse problems: multiple intracranial configurations can produce similar scalp observations, and introducing a latent variable provides a natural probabilistic mechanism to represent this ambiguity rather than forcing a single deterministic reconstruction~\cite{asadzadeh2020systematic,zorzos2021advances}.

This perspective also helps contextualize the regional variability discussed above. For deeper or more structurally complex targets (e.g., the anterior hippocampus), scalp observability is expected to be reduced, and thus a larger portion of iEEG variability may remain underconstrained by EEG~\cite{piastra2021comprehensive,afnan2024eeg,vorwerk2024global}. In NeuroFlowNet, such underconstrained components may be represented through the latent pathway $Z$, enabling the model to avoid collapsing to an ``average'' waveform and instead generate diverse yet EEG-consistent iEEG samples~\cite{papamakarios2021normalizing}. We emphasize that the above points constitute a physiologically motivated interpretation of $Z$ rather than an established one-to-one identification of latent dimensions with specific neural generators or physiological processes. Future work should test these hypotheses using dedicated interpretability analyses (e.g., linking latent factors to band-limited power modulation, event-related signatures, or unit/spike-informed markers) on larger cohorts with broader brain-state coverage~\cite{boran2020dataset}.

\subsection{Discussion of Clinical Translation and Robustness Considerations}
NeuroFlowNet demonstrates the feasibility of reconstructing MTL iEEG patterns from scalp EEG using a public synchronized EEG-iEEG dataset. The cross-session evaluation further provides a more stringent within-subject robustness test than random trial splitting, as the model is required to generalize to entire recording sessions excluded from training. The broadly comparable performance between cross-session and cross-trial evaluations suggests that NeuroFlowNet is not merely exploiting trial-level redundancy within individual sessions, but instead captures a relatively stable, subject-specific EEG-to-iEEG mapping.

Nevertheless, the present evidence should be interpreted within a subject-specific and task-dependent scope~\cite{boran2020dataset}. The dataset consists of recordings from epilepsy patients with individualized implantation schemes, and intracranial electrode coverage is not standardized across subjects~\cite{jobst2020intracranial}. Consequently, direct cross-subject generalization cannot be rigorously established using the current data. Thus, although our findings support the methodological potential of non-invasive iEEG reconstruction, they do not yet warrant strong clinical claims, particularly for high-stakes applications such as non-invasive pre-surgical localization~\cite{jobst2020intracranial}.

Clinical translation of EEG-to-iEEG reconstruction ultimately depends on robust performance across both subjects and brain states. Inter-subject variability arises from differences in anatomy, skull conductivity, pathology, medication, and electrode implantation targets, all of which can alter the scalp-to-intracranial mapping~\cite{vorwerk2024global,asadzadeh2020systematic,zorzos2021advances,wang2026pa}. Cross-state variability is also expected, as intracranial dynamics change substantially across vigilance and cognitive conditions (e.g., rest vs. task engagement), and across clinically relevant contexts (e.g., interictal vs. ictal periods)~\cite{pershin2023vigilance,jobst2020intracranial}. These factors may introduce distribution shifts that challenge purely subject-specific modeling and motivate broader validation~\cite{piastra2021comprehensive,afnan2024eeg}.

To move toward clinical readiness, several steps are necessary. First, larger multi-subject datasets with more comparable intracranial coverage (or standardized ROI definitions) are required to enable systematic evaluation of cross-subject generalization and the development of adaptation strategies that account for anatomical and physiological differences~\cite{jobst2020intracranial,vorwerk2024global}. Second, robustness should be assessed across heterogeneous brain states, ideally including resting state, sleep, multiple cognitive tasks, and interictal/ictal segments, to establish state generalization~\cite{pershin2023vigilance,jobst2020intracranial}. Third, the clinical value of reconstructed signals should be validated in downstream endpoints where performance can be directly linked to decision-making, such as interictal spike detection, seizure onset zone (SOZ) localization, or surgical outcome prediction~\cite{jobst2020intracranial}. Taken together, we position NeuroFlowNet as a methodological step toward non-invasive deep-brain signal inference, while recognizing that comprehensive cross-subject and cross-state validation is an essential prerequisite for clinical translation~\cite{jobst2020intracranial}.

\section{Conclusion}
In this study, we introduced NeuroFlowNet, a novel generative framework based on conditional normalizing flows, to address the enduring challenge of generating high-fidelity (band-limited) iEEG from non-invasive EEG within the preserved bandwidth. Our work makes a pivotal contribution by demonstrating, for the first time, the feasibility of reconstructing iEEG signals across multiple MTL subregions. By directly modeling the conditional probability distribution, NeuroFlowNet provides a latent-variable mechanism for representing residual iEEG variability that is not fully determined by scalp EEG. Our results confirm that the generated signals not only achieve high fidelity in temporal waveforms and spectral characteristics but also successfully preserve the complex inter-channel correlation structure, reflecting the underlying functional connectivity of the MTL network. The cross-session experiments further show that these properties remain broadly stable when evaluation is shifted from unseen trials to unseen recording sessions from the same subject, strengthening the robustness claim of the method. In addition, a baseline comparison against representative deterministic regressors shows that NeuroFlowNet yields lower errors in the inter-channel correlation matrix under the same within-subject protocol, supporting its advantage in structure-preserving reconstruction rather than purely pointwise fitting. Conditional sampling with 20 latent draws per EEG segment further demonstrates non-degenerate stochastic reconstructions and a strong positive association between predictive variance and reconstruction error, providing direct evidence of generative behavior under the present protocol. The latency analysis further shows that the final model can generate a 1-s iEEG segment in only a few milliseconds on a modern GPU, supporting its practical potential for real-time or near-real-time monitoring. These findings suggest that probabilistic iEEG reconstruction may provide a useful non-invasive window into deep brain dynamics, particularly for analyses that require both waveform plausibility and multi-channel dependency structure.

However, we acknowledge certain limitations. The model's performance exhibits variability across different MTL subregions, with deeper and more structurally complex areas like the anterior hippocampus proving more challenging to reconstruct accurately. This suggests that the signal-to-noise ratio of the conditioning EEG and the inherent complexity of the target signal are key performance determinants. Additionally, since iEEG is downsampled to 200~Hz with anti-aliasing, the present results and spectral claims are limited to the preserved low-/mid-frequency range ($\leq$100~Hz; evaluated mainly in 0.5-50~Hz and 8-13~Hz), and do not address high-frequency intracranial biomarkers such as high-gamma/HFO. The conditional sampling analysis also showed under-coverage of empirical 90\% prediction intervals, indicating that the current uncertainty estimates are informative but not fully calibrated. Moreover, the present study does not yet establish cross-subject or cross-state generalization, and it does not include a protocol-matched comparison with alternative generative models such as diffusion models or GANs. Clinical deployment should therefore be considered premature until the framework is validated on larger cohorts, heterogeneous brain states, and broader generative baselines.

Future work will focus on four main directions: 1) Enhancing the model architecture by incorporating more sophisticated conditioning mechanisms that can better leverage spatial information from the EEG montage. 2) Expanding the framework to a subject-independent model by collecting larger-scale datasets with more standardized and comparable electrode implantation coverage across subjects, which will enable a systematic evaluation of cross-subject generalization and support the development of advanced domain adaptation techniques to account for inter-subject anatomical and physiological variability~\cite{zhao2026rmetnet,wang2026pa,wang2026cfspmnet}. 3) Improving uncertainty calibration and comparing NeuroFlowNet with protocol-matched generative baselines, including diffusion- or GAN-based conditional generators. 4) Validating the clinical utility of the generated iEEG in downstream tasks, such as automated seizure detection and cognitive state decoding, to bridge the gap between technical innovation and practical application.

\section*{CRediT authorship contribution statement}
\textbf{Dongyi He}: Conceptualization, Methodology, Software, Data curation, Writing - original draft, Writing - review \& editing, Visualization.
\textbf{Bin Jiang}: Conceptualization, Supervision, Writing - review \& editing, Funding acquisition, Writing - original draft, Formal analysis.
\textbf{Kecheng Feng}: Writing - review \& editing, Visualization, Formal analysis.
\textbf{Luyin Zhang}: Writing - review \& editing.
\textbf{Ling Liu}: Writing - review \& editing.
\textbf{Yuxuan Li}: Writing - review \& editing.
\textbf{Yun Zhao}: Funding acquisition, Formal analysis, Writing - review \& editing, Supervision, Resources
\textbf{He Yan}: Supervision, Resources
\textbf{All authors have read and approved the final manuscript.}

\section*{Declaration of competing interest}
The authors declare that they have no known competing financial interests or personal relationships that could have appeared to influence the work reported in this paper.

\section*{Ethics statement}
This study was conducted in publicly available datasets, thus ethical approval was not required.

\section*{Declaration of generative AI and AI-assisted technologies in the writing process.}
During the preparation of this work the author(s) used ChatGPT in order to check the grammar and spelling of the manuscript. After using this tool/service, the author(s) reviewed and edited the content as needed and take(s) full responsibility for the content of the published article.

\section*{Funding Declaration}
This research was funded by the Scientific and Technological Research Program of the Chongqing Education Commission (KJZD-K202303103, KJQN202501104) and Chongqing Municipal Key Project for Technology Innovation and Application Development (CSTB2024TIAD-KPX0042).

\bibliographystyle{elsarticle-num} 
\bibliography{references}

\end{document}